\documentclass[12pt,letterpaper,article,superscriptaddress]{revtex4}

\usepackage{natbib}

\usepackage[dvips]{graphicx}
\usepackage{epsfig}
\usepackage{dcolumn}
\usepackage{amssymb,amsfonts,amsmath,amsthm}
\usepackage[tight,hang,raggedright,normalsize]{subfigure}

\usepackage{setspace}

\begin{document}
\pagestyle{empty}

\begin{center}
\vspace{1cm}
\hspace{9cm}
{JLAB-PHY-09-993}
\end{center}

\title{\Large\bf Theory Support for the Excited Baryon Program at the JLab 12 GeV
Upgrade\footnote{Notice: Authored by
Jefferson Science Associates, LLC under U.S. DOE Contract No. DE-AC05-06OR23177. 
The U.S. Government retains a non-exclusive, paid-up, irrevocable, world-wide
license to publish or reproduce this manuscript for U.S. Government purposes.
}\\
\vspace{0.5cm}
}

\vspace{1cm}

%%%%%%%%%%%%%%%%%%%% Authors %%%%%%%%%%%%%%%%%%%%%
\author{I. Aznauryan}
\affiliation{ Thomas Jefferson National
Accelerator Facility, Newport News, VA 23606, USA}
\affiliation{ Yerevan Physics Institute, 375036 Yerevan, Armenia}
\author{V. Braun}
\affiliation{Institute for Theoretical Physics, University of Regensburg, 93040,Regengurg, Germany}
\author{ V. Burkert}
\affiliation{ Thomas Jefferson National
Accelerator Facility, Newport News, VA 23606, USA}
\author{S. Capstick}
\affiliation{Department of Physics, Florida State University, USA}
\author{ R. Edwards}
\affiliation{ Thomas Jefferson National
Accelerator Facility, Newport News, VA 23606, USA}
\author{I.C.Cloet}
\affiliation{ Department of Physics, University of Washington, Seattle, WA 98195, USA}
\author{M. Giannini}
\affiliation{University of Genova and National Institute of Nuclear Physics.
Genova, Via Dodecaneso, 33, Italy}
\author{T.-S. H. Lee}
\affiliation{ Thomas Jefferson National
Accelerator Facility, Newport News, VA 23606, USA}
\affiliation{Physics Division, Argonne National Laboratory, Argonne, 
IL 60439, USA}
\author{ H.-W. Lin}
\affiliation{ Thomas Jefferson National
Accelerator Facility, Newport News, VA 23606, USA}
\author{ V. Mokeev}
\affiliation{ Thomas Jefferson National
Accelerator Facility, Newport News, VA 23606, USA}
\affiliation{Skobeltsyn Nuclear Physics Institute at Moscow State University,
Moscow 119899, Leninskie gory, OEPVAYa, Russia}
%\author{M.V. Polyakov}
%\affiliation{Ruhr Univ. Bochum, Inst. Theor. Physik II, Bochum, D-44801, Germany}
%\affiliation{Petersburg Nuclear Physics Institute, Gatchina, Leningrad District,
%188300 St.Petersburg, Russia}
\author{C.D. Roberts}
\affiliation{Physics Division, Argonne National Laboratory, Argonne, IL 60439, USA}
\author{E. Santopinto}
\affiliation{ National Institute of Nuclear Physics.
Genova, Via Dodecaneso, 33, Italy}
\author{P. Stoler}
\affiliation{Physics Department, Rensselaer Polytechnic Institute, 
Troy, NY 12180, USA}
\author{Q. Zhao}
\affiliation{Institute of High Energy Physics,
Chinese Academy of Sciences, Beijing 100049, P.R. China}
\author{B.S. Zou}
\affiliation{Institute of High Energy Physics,
Chinese Academy of Sciences, Beijing 100049, P.R. China}

%\begin{abstract}
%\end{abstract}

\maketitle

\newpage
\setcounter{page}{0}
%\makeatletter
%\renewcommand\l@section[2]{%
%  \ifnum \c@tocdepth >\z@
%    \addpenalty\@secpenalty
%    \addvspace{1.0em \@plus\p@}%
%    \setlength\@tempdima{2.5em}%
%    \begingroup
%      \parindent \z@ \rightskip \@pnumwidth
%      \parfillskip -\@pnumwidth
%      \leavevmode \bfseries
%      \advance\leftskip\@tempdima
%      \hskip -\leftskip
%      #1\nobreak\ 
%      \leaders\hbox{$.$}
%     \hfil \nobreak\hb@xt@\@pnumwidth{\hss #2}\par
%    \endgroup
%  \fi}
%\makeatother
%\newpage
\tableofcontents
\pagestyle{plain}

\newpage
\section*{Abstract}
This document outlines major directions in theoretical support for the measurement
of nucleon resonance transition form factors at the JLab 12 GeV upgrade with the
CLAS12 detector. Using single and double meson production, prominent resonances in
the mass range up to 2 GeV will be studied in the range of photon virtuality $Q^2$
up to 12~GeV$^2$ where quark degrees of freedom are expected to dominate. High
level theoretical analysis of these data will open up opportunities to  understand
how the interactions of dressed quarks create the ground and excited nucleon states
and how these interactions emerge from QCD.  The paper reviews the current
status and the prospects of QCD based model approaches that relate 
phenomenological information on transition form factors to the non-perturbative strong interaction
mechanisms, that are responsible for resonance formation.

%\begin{center}
%{\Large Contents}
%\end{center}

%\renewcommand{\baselinestretch}{0.5}

\newpage
\section{Introduction} 

%\documentclass[preprint,aps,tightenlines,showpacs,superscriptaddress]{revtex4}
%#%\documentclass[prl,eqsecnum,twocolumn,floats,aps,showpacs,superscriptaddress]{revtex4}
%\usepackage[dvips]{graphicx}
%\usepackage{dcolumn}
%\usepackage{bm}
%\usepackage{epsfig}

%\usepackage{showlabels}
%\renewcommand{\case}{\frac}

\newcommand{\sla}[1]{\not\! #1}
\def\ohalf{{\textstyle{1\over 2}}}
\def\half{{\textstyle{1\over 2}}}
\def\vqhalf{{\textstyle{\vec{Q}\over 2}}}
\def\qhalf{{\textstyle{Q\over 2}}}
\def\osix{{\textstyle{1\over 6}}}
\def\vqsix{{\textstyle{\vec{Q}\over 6}}}
\def\thalf{{\textstyle{3\over 2}}}
\def\fourth{{\textstyle{1\over 4}}}
\def\tfor{{\textstyle{3\over 4}}}

\newcommand\Tr{\,{\rm Tr}\,}
\newcommand\re{\Re\mbox{e}}
\newcommand\im{\Im\mbox{m}}
\newcommand{\slas}[1]{\not\! #1}
\newcommand{\beq}{\begin{equation}}
\newcommand{\eeq}{\end{equation}}

%%%\begin{document}

%%%\section{Introduction}
Nucleons and baryons in general, have played an essential role in the 
development of our understanding of the strong interaction. The concept 
of quarks was first made manifest through the study of baryon
spectroscopy, which subsequently led to the development of dynamical constituent quark models
(CQMs) in the late 1960's \cite{cko69} and further developed in the 
1970's \cite{isgur,morp}. As a result of intense experimental 
and theoretical effort, especially in recent years, it has become clear 
that the structure of the nucleon and its excited states ($\Delta^*$ and $N^*$) is much
more complex than what can be described in terms of constituent quarks only. 
The structure of low-lying baryon states, as revealed by electromagnetic 
probes at low momentum transfer, can be described reasonably well by 
adding meson-baryon effects phenomenologically to the predictions from 
constituent quark models 
\cite{Cloet:2002eg,Az93,cqm-rel-1,cqm-rel-2,cqm-rel-3,cqm-rel-31,cqm-rel-4}. However, a 
fundamental understanding of the 
properties of the nucleon and its excited states at short distances, 
which are accessible using probes with sufficiently high momentum
transfer, demands the full machinery of Quantum Chromodynamics (QCD). In
recent years, there has been tremendous progress in this direction. 
Constituent quark models have been greatly refined by using fully 
relativistic treatments~\cite{Az93,cqm-rel-1,cqm-rel-2,cqm-rel-4} and by 
including sea quark components~\cite{cqm-seaq,cqm-seaq1,cqm-seaq2}. The hypercentric CQM with improved
treatment of constituent quark interactions ~\cite{cqm-rel-3,cqm-rel-31} has emerged. 
A covariant model based 
on the Dyson-Schwinger equations~\cite{dsm-1} (DSE) of QCD is now emerging 
as a well-tested and well-constrained tool to interpret baryon data 
directly in terms of current quarks and gluons. This approach also provides
a link between the phenomenology of dressed current quarks and Lattice QCD (LQCD).  
Relations between baryon transition form factors and the Generalized Parton 
Distributions (GPDs) have also been formulated~\cite{gpd-1,gpd-2} that connect
these two different approaches to describing baryon structure. On a 
fundamental level, Lattice QCD is progressing rapidly in making contact 
with the excited baryon data. The USQCD Collaboration, involving JLab's LQCD group, 
has been formed to perform calculations for predicting the baryon spectrum 
and $\gamma_vNN^*$ transition form factors.

%%%%%%%%%%%%%%%%%%%%%%%%%%%%%%%%%%%%%%%%%%%%%%%%%%%%%%%%%%%%%%%%%%%%%%%%%%% 
%\begin{figure}[ht]
%\centering
%\includegraphics[clip,width=7.6cm]{INTRODUCTION/bruno-fig1.eps}
%\includegraphics[clip,width=8cm]{INTRODUCTION/F12-pR-sim2.eps}

%\caption{Lattice QCD calculations of transition form factors. Left panel:
%$N$-$\Delta(1232)$ transition form factors $G_M$, $G_E$, and $G_C$ vs
%$Q^2$. Empirical values (solid squares) are extracted by EBAC from world 
%data within a dynamical model. The LQCD results are from 
%Ref.~\cite{lqcd-dinna}. Right panel: $F_{1}(Q^2)$ and $F_{2}(Q^2)$ for the 
%$N$-$P_{11}(1440)$ transition. Empirical values are from the CLAS 
%Collaboration \cite{Az08,Mo08m} and the LQCD results are from Ref.~\cite{lqcd-jlab}.}
%\label{fig:lattice-result}

%\end{figure}
%%%%%%%%%%%%%%%%%%%%%%%%%%%%%%%%%%%%%%%%%%%%%%%%%%%%%%%%%%%%%%%%%%%%%%%%%%% 
 
On the experimental side, extensive data on electromagnetic meson 
production have been obtained at JLab, MIT-Bates, LEGS, MAMI, ELSA, and 
GRAAL in the past decade. The analyses of these data and the data 
expected in the next few years before the start of experiments with the
JLab 12-GeV upgrade, will resolve some long-standing problems in baryon 
spectroscopy and will provide new information on the structure of $N^*$ 
states. To enhance this effort, the Excited Baryon Analysis Center (EBAC) 
was established in 2006 and is now making rapid progress in this direction.
Analysis models developed at Mainz, JLab, GWU, and Bonn are also being
greately refined to analyze the recent data. 
Significant progress from this experiment-theory joint effort has been made 
in the past few years. 

With the 12 GeV upgrade of CEBAF at JLab and the
development of experimental facilities
 at Mainz and Bonn,
new opportunities for investigating the spectrum and structure of excited baryon states will soon
become available. To develop research programs for this new era, 
a workshop on 
Electromagnetic $\gamma_vNN^*$ Transition Form Factors
was held at Jefferson Laboratory, October 13-15, 2008 \cite{emnns}.
The main objectives of the workshop were
(a) to review the status of the $\gamma_vNN^*$ transition form factors extracted from the meson
electroproduction data, and (b) to call for the theoretical interpretations of 
the extracted $N$-$N^*$ transition form factors, that enable access to the mechanisms 
responsible for the N* formation and to their emergence from QCD.

This document summarizes the contributions of workshop
participants that provide theoretical support for the excited baryon program at the 12 GeV
energy upgrade at JLab.

%%\end{document}

%\newpage
\section{ Physics from Lattice QCD} 
  %\documentclass[12pt]{article}
%\setlength{\textwidth}{16cm}
%\setlength{\textheight}{22.5cm}
%\setlength{\oddsidemargin}{0.25cm}
%\setlength{\topmargin}{0.25cm}

% Somewhat wider and taller page than in art12.sty
%\topmargin -0.4in  \headsep 0.0in  \textheight 9.0in
%\oddsidemargin -0.8in  \evensidemargin -0.8in  \textwidth 7.8in

%\footnotesep 14pt
%\floatsep 28pt plus 2pt minus 4pt      % Nominal is double what is in art12.sty
%\textfloatsep 40pt plus 2pt minus 4pt
%\intextsep 28pt plus 4pt minus 4pt

%\usepackage{amssymb}
%\usepackage{epsfig}
%\usepackage{graphicx}

%\begin{document}

%\title{\vspace*{-1in}$N^\ast$ Physics from Lattice QCD}
%\author{Robert Edwards and Huey-Wen Lin}
%\date{November 18, 2008}
%\maketitle

%\begin{center}
%{\large Robert Edwards and Huey-Wen Lin}
%\\
%\end{center}

Quantum Chromodynamics (QCD), when combined with the electroweak
interactions, underlies all of nuclear physics, from the spectrum and
structure of hadrons to the most complex nuclear reactions. Lattice
gauge calculations enable the \textit{ab initio} study of many of the
low-energy properties of QCD. There are significant efforts underway
internationally to use lattice QCD to directly compute properties of
the ground and excited state nucleon and, generically, the baryon
spectrum of matter, including spectrum and structure.

The Hadron Spectrum Collaboration involving the Lattice Group at
Jefferson Lab, Carnegie Mellon University, Univ. of Maryland, and
Trinity College (Dublin) has embarked on an ambitious program to
compute the high lying excited state spectrum of baryons and mesons,
as well as their (excited state) electromagnetic transition
form-factors up to $Q^2 \sim 10~{\rm GeV}^2$. A particularly important
quantity to compute is the photo-coupling value for exotic mesons
which is of relevance for experiments in the future Hall D at
JLab. With the new techniques that will be used to extract resonance
information, it is intended that the spectrum and couplings that are
determined can be used to provide valuable comparisons with
experimental data, and provide input for programs like EBAC.

There are several key technologies needed in this campaign. To
adequately resolve excited state energies and to keep the
calculational costs manageable, an anisotropic lattice formulation is
used with three flavors of quarks - two light and a strange
quark. These new type of lattices require a significant amount of
computing resources since previous lattice configurations cannot be
(re)used. As described in Ref.~\cite{Lin:2008pr}, a successful program
is underway to generate these lattices using DOE and NSF computing
resources, and those available within the USQCD collaboration,
including clusters at JLab. It is anticipated that the production of
configurations at the {\em physical} pion mass will proceed early in 2009
using the next generation of Cray supercomputers at ORNL.

\begin{figure}[!hbt]
\begin{center}
\begin{tabular}{lc}
\includegraphics[width=0.52\textwidth,clip=true]{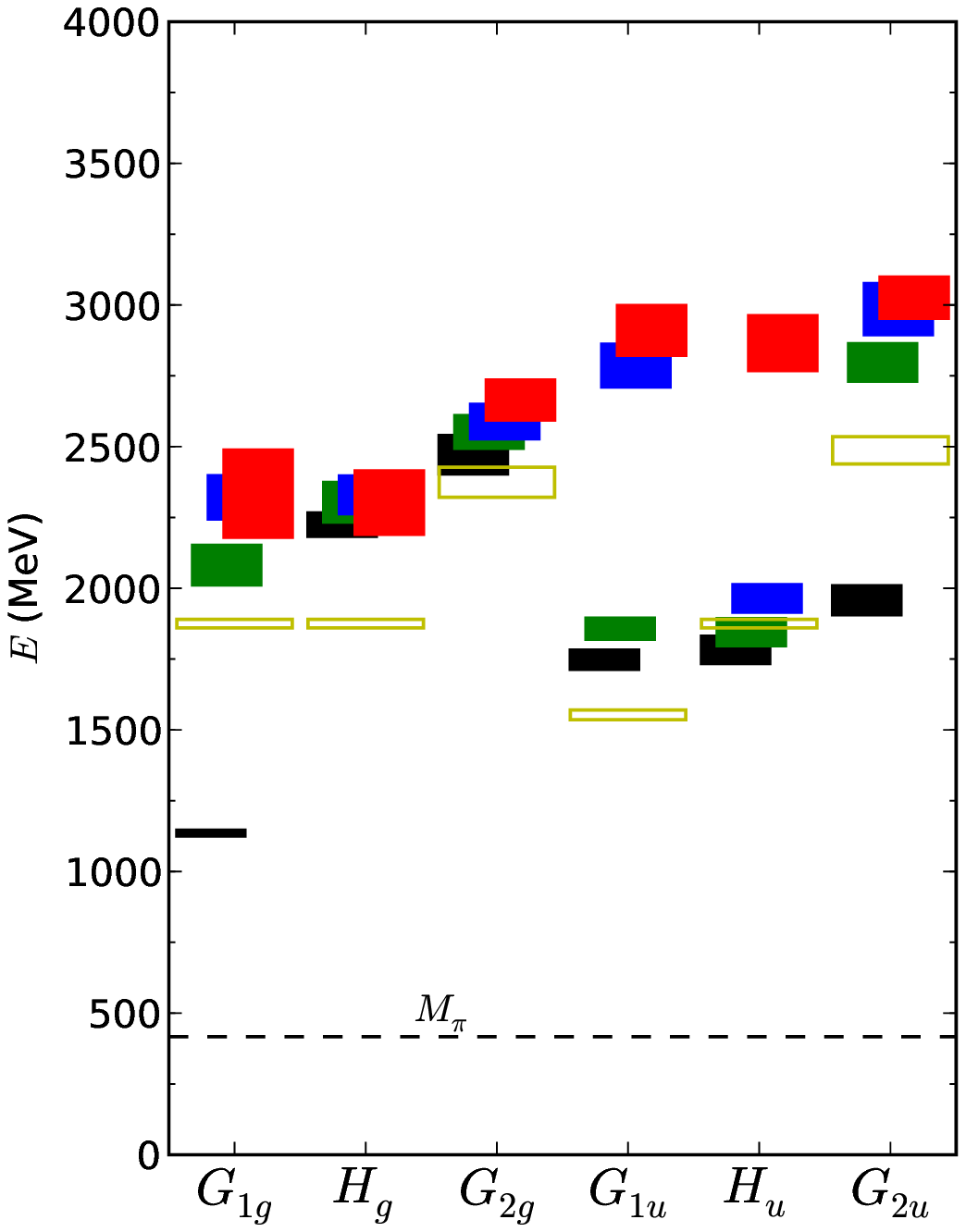} &
\includegraphics[width=0.5\textwidth,clip=true]{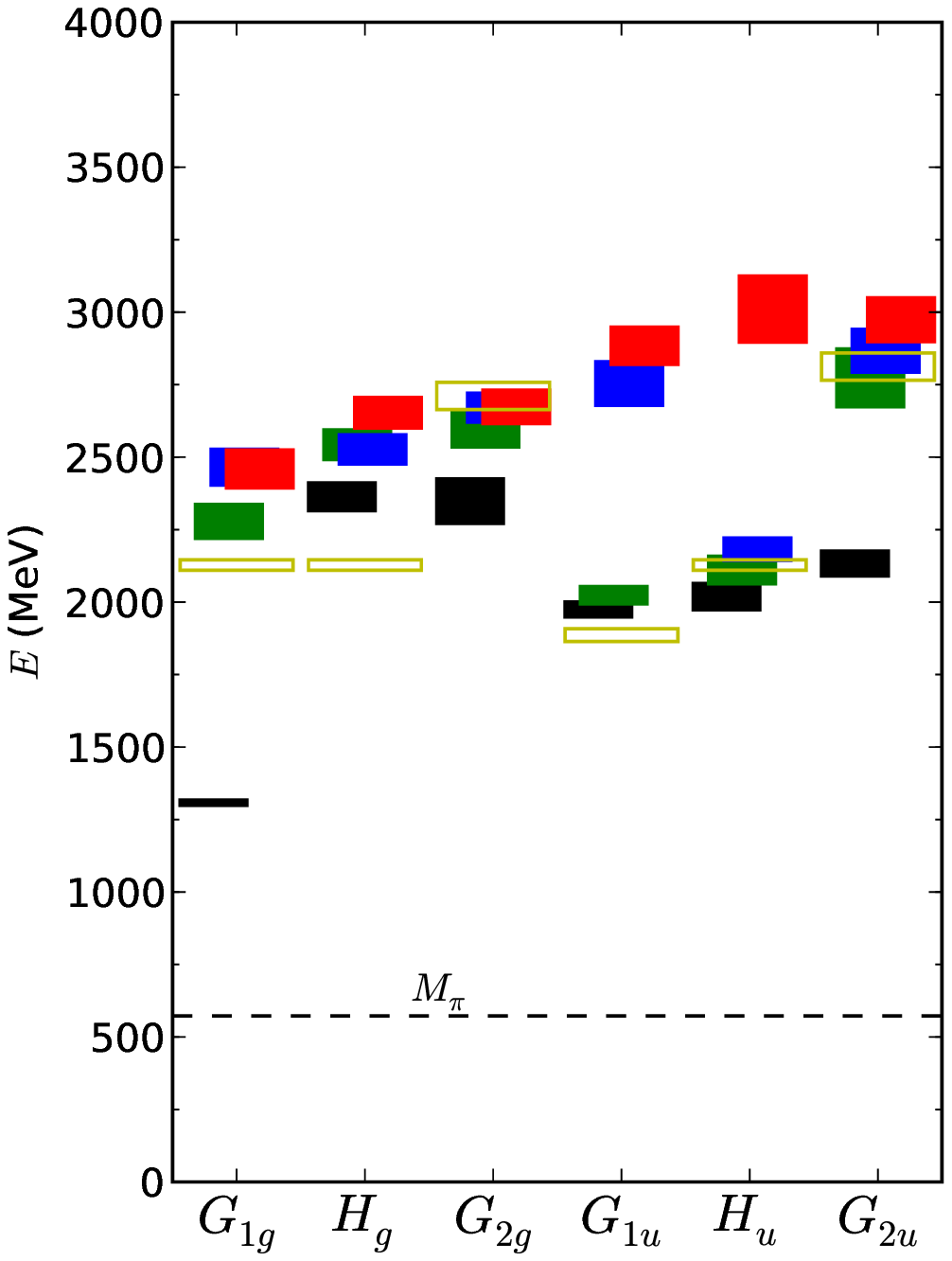}
%\includegraphics[width=0.6\textwidth,clip=true]{LQCD-1/400mev_pion_with_thresholds.eps} &
%\hspace{-1in}
%\includegraphics[width=0.6\textwidth,clip=true]{LQCD-1/572mev_pion_with_thresholds.eps}
\end{tabular}
\end{center}
\caption{The energies obtained for each symmetry channel of
isospin $\frac{1}{2}$ baryons are shown based on the
$2.64fm^3$ $N_f=2$ lattice QCD data for $m_{\pi}$ = 400 
MeV (left panel) and $m_{\pi}$ = 572 MeV (right panel). 
The scale shows energies in Mev and errors are indicated by the vertical
size of the box. The gold open boxes show $N\pi$ threshold states.} \label{fig:boxplot_m400}
\end{figure}

Another key component in the hadron spectrum campaign is the use of
variational techniques for constructing correlators. The hadron
creation operators used in the correlators should have significant
overlap with the hadron states of interest. In
Ref.~\cite{Basak:2005aq}, group theoretical techniques have been used
to construct non-local interpolating fields that characterize possible
hadron states. Their spins are classified according to irreducible
representations of the cubic rotation group -- the remnants of the
rest frame Lorentz group when discretized. These large bases of
operators are used in a variational calculation which allows for the
extraction of a large number of excited states. In
Figure~\ref{fig:boxplot_m400} is shown the extracted energies of
highly excited levels of the nucleon spectrum at unphysical pion
masses using two flavors of anisotropic quarks \cite{Bul:09}.  The technique to
reconstruct the continuum spin states which are broken into lattice
irreducible representations has been developed in
Ref.~\cite{Dudek:2007wv}. These techniques are being used now in
light quark mass calculations of the baryon spectrum as well as the
meson spectrum over the $N_f=2+1$ configurations.

%%%%%%%%%%%%%%%
%%%%%%%%%%%%%%%%%

\begin{figure}[!hbt]
\begin{center}
\includegraphics[width=0.47\textwidth]{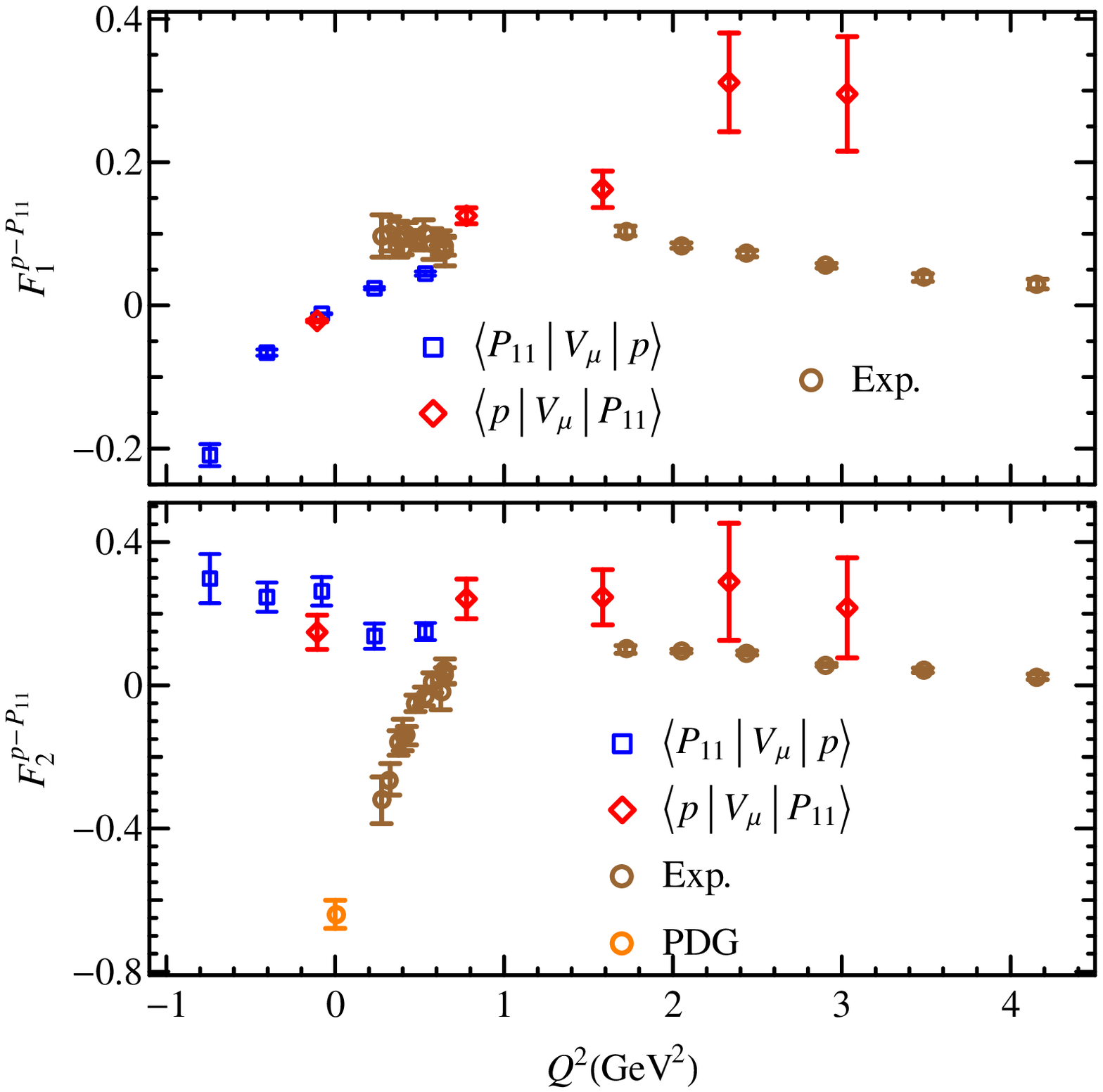}
\includegraphics[width=0.47\textwidth]{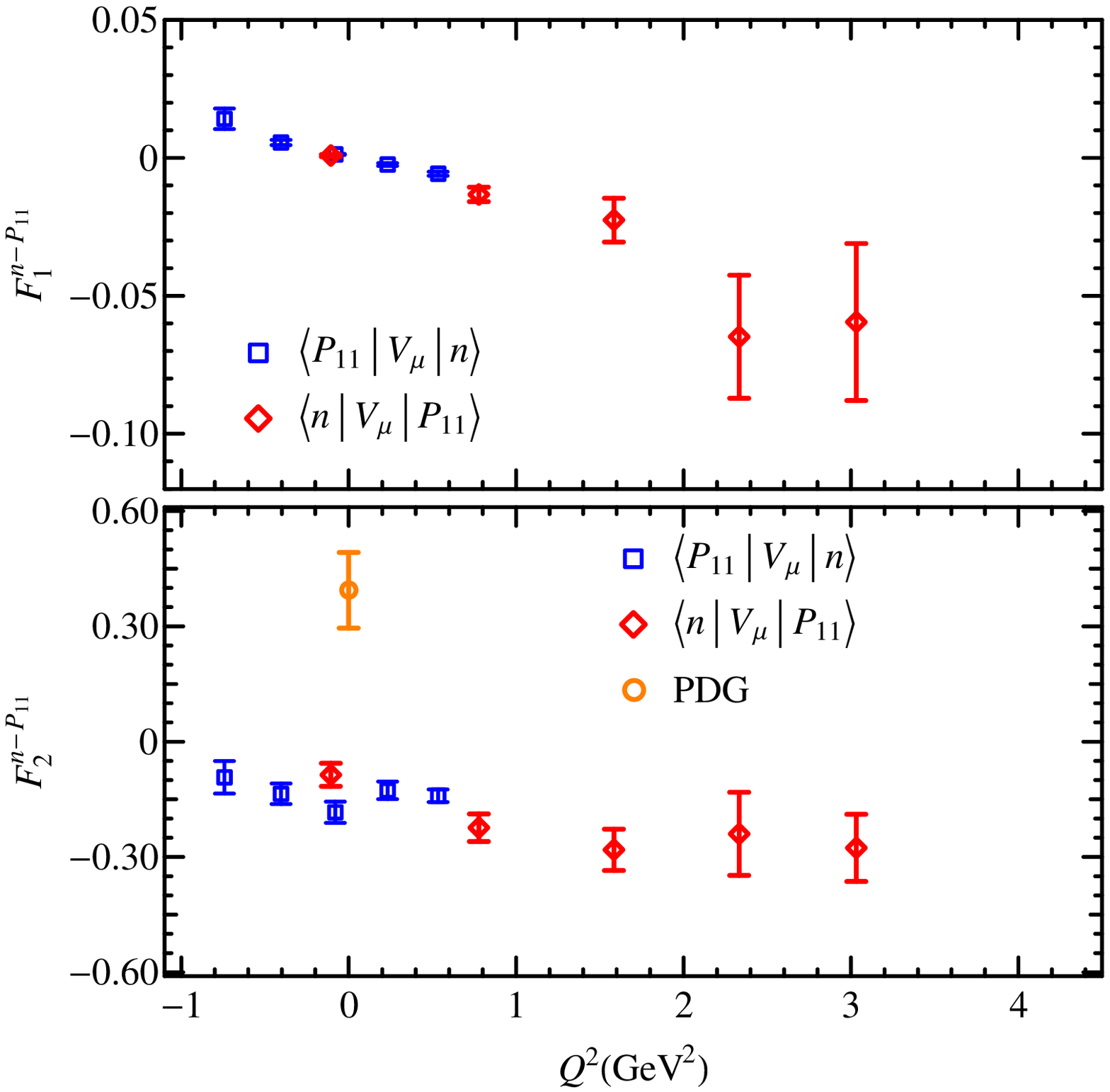}
\end{center}
\caption{\label{fig:F-nucleon}Left panel: proton-Roper form factors $F_{1,2}^*$ obtained from CLAS experiments
%\cite{Mokeev:2006an,Joo:2005gs,Burkert:2004sk,Burkert:2002nr} 
and PDG
%\cite{PDBook} 
number (circles) and lattice methods (squares, diamonds).
Right panel: neutron-Roper form factors $F_{1,2}^*$ obtained from PDG
%\cite{PDBook} 
number (circles) and lattice methods (squares, diamonds)}
\end{figure}

At large enough quark mass, the ground and probably many of the
excited baryon states are stable under the strong
interactions. However, as the quark mass decreases, decay channels
open up which are of lower energy compared to the state of interest. A
critically key component in this hadron spectrum campaign is
identifying how these single particle and multi-particle states shift
with changes in the physical volume of the lattice. These finite
volume techniques, developed by L\"uscher, have successfully been used
in mesonic systems, but their use in baryonic systems is relatively
new and are actively under investigation. 

%An important tool is the use
%of multi-hadron operators to give additional overlap onto multi-particle
%states. Using the spin identification techniques from
%Ref.~\cite{Dudek:2007wv} in these multi-hadron operators provides
%the lattice analog of partial wave analysis methods 
%and aides in state
%indentification.

From the excited energies of the nucleon spectrum, one can compute
electromagnetic form-factors. First exploratory results have been
obtained in Ref.~\cite{Lin:2008qv} for the excited nucleon
$P_{11} - N$ transition using a very simple basis of
operators. The main result is shown in Figure~\ref{fig:F-nucleon}.
The low $Q^2$ region for $F_2(Q^2)$, at these very large unphysical
pion masses shows large deviations from experiment, consistent with
many statements that the pion cloud effects are stronger in excited
state systems compared to the ground states.  However, these first
preliminary results are very encouraging given the very limited
operator basis.  Work is underway now using the previously developed
full basis of nucleon operators for a more accurate computation of the
excited nucleon form-factors at much smaller pion masses using the
$N_f=2+1$ configurations already produced. In addition, the ground and
excited state hyperon transition form-factors will also be
computed. It is not clear what kind of statistical accuracy that might
be achieved - it is very sensitive to the system of interest, what
excited level, and what $Q^2$ (many are available in one calculation). 
The results in Fig.~\ref{fig:F-nucleon} are illustrative.

The $Q^2$ range in these current form-factor calculations is typically
up to about $3$ to $4~{\rm GeV}^2$. To go to about $10~{\rm GeV}^2$
requires some different techniques. One method is to go to smaller
lattice spacing $a$ where $Q^2\sim 1/a^2$. Since more than one lattice
spacing is needed for continuum extrapolations, this change will
happen, probably in late 2009. However, a more immediate method
involves going to the (anti)-Breit frame between the initial and final
nucleon states, whereby the $Q^2$ is maximized. Using this technique
(Ref.~\cite{Lin:2008gv}), the previous calculation for the $P_{11} - N$ 
transition was extended up to $6~{\rm GeV}^2$, again at unphysically
large pion masses, and it seems feasible to go somewhat higher
$Q^2$. This (anti)-Breit frame technique is being used in lighter pion
mass calculations. As the pion mass decreases, the previous results at
time-like $Q^2$ will slide to larger (positive) values, and hence the
$Q^2_{max}$ values will also increase to greater than $7~{\rm GeV}^2$.
Again, the obtainable statistical error in this large $Q^2$ region is
not known at this time.

In parallel with this work of the computation of the excited nucleon
spectrum, significant effort is going into the calculation of the
excited meson spectrum.  First efforts (Ref.~\cite{Dudek:2007wv}) have
gone into an extensive calculation of the excited charmonium
spectrum. 
%Current work in this same system has shown that the
%calculation of the fourth excited state radiative transition
%$\psi^{(iv)}\rightarrow\eta_c\gamma$ is possible. 
The goal of this work is the determination of the charmonium version
of the $1^{-+}$ photo-coupling as phenomenological input for
GlueX. Working is proceeding now on the calculation of the $1^{-+}$
photocoupling at the strange quark mass scale, and soon for the light
quark mass scale. 

%In addition to the determination of excited state resonance masses and
%electromagnetic transition form-factors, the calculation of the
%$N^\as$ distribution amplitudes provides valuable phenomenological
%information.

%\end{document}

%\newpage
\section{ Charting the interaction between light quarks }

\newcommand{\lsim}{\mathrel{\rlap{\lower4pt\hbox{\hskip0pt$\sim$}} 
\raise1pt\hbox{$<$}}}           %less than or approx. symbol 
\newcommand{\gsim}{\mathrel{\rlap{\lower4pt\hbox{\hskip0pt$\sim$}} 
\raise1pt\hbox{$>$}}}           %greater than or approx. symbol 

%\setlength{\parindent}{0pt}
%\setlength{\parskip}{0.5\baselineskip}

%\begin{document}

%\begin{center}
%\Large{\textbf{NSTAR12: Charting the interaction between light quarks}}
%\\[2ex]

%\large{Ian C.\ Clo\"et}
%\\
%\textit{Department of Physics, University of Washington, Seattle WA 98195, USA}
%\\[1ex]

%\large{Craig D.\ Roberts}
%\\
%\textit{Physics Division, Argonne National Laboratory, Argonne IL 60439, USA}
%\end{center}

%\begin{center}
%%{\bf NSTAR12: Charting the interaction between light quarks}
%%\\
%%\vspace{0.5cm}
%{\large Ian C.\ Clo\"et and  Craig D.\ Roberts}
%\\
%\end{center}

%{\large Ian C.\ Clo\"et}
%\\
%{\it Department of Physics, University of Washington, Seattle WA 98195, USA}
%\\
%{\large Craig D.\ Roberts}
%\\
%{\it Physics Division, Argonne National Laboratory, Argonne IL 60439, USA}
%\end{center}

Two of the basic motivations for an upgraded JLab facility are the needs: to determine the essential nature of light-quark confinement and dynamical chiral symmetry breaking (DCSB); and to understand nucleon structure and spectroscopy in terms of QCD's elementary degrees of freedom.  In addressing these questions one is confronted with the challenge of elucidating the role of quarks and gluons in hadrons and nuclei.   In accepting that challenge one steps immediately into the domain of relativistic quantum field theory where within the key phenomena can only be understood via nonperturbative methods.  

It is a fundamental fact that the physics of hadrons is dominated by two \emph{emergent phenomena}: confinement; namely, the empirical truth that quarks have not hitherto been detected in isolation; and DCSB, which is responsible, amongst many other things, for the large mass splitting between parity partners in the spectrum of light-quark hadrons, even though the relevant current-quark masses are small.  Neither of these phenomena is apparent in QCD's Lagrangian and yet they play a principal role in determining the observable characteristics of real-world QCD.  

In connection with confinement it is worth emphasizing at the outset that the potential between infinitely-heavy quarks measured in numerical simulations of quenched lattice-regularised QCD -- the so-called static potential -- is simply not relevant to the question of light-quark confinement.  In fact, it is quite likely a basic feature of QCD that a quantum mechanical potential between light-quarks is impossible to speak of because particle creation and annihilation effects are essentially nonperturbative.  A perspective on confinement was laid out in Ref.\,\cite{Krein:1990sf}.  Expressed simply, confinement can be related to the analytic properties of QCD's Schwinger functions, which are often loosely called Euclidean-space Green functions.  For example, it can be read from the reconstruction theorem that the only Schwinger functions which can be associated with expectation values in the Hilbert space of observables; namely, the set of measurable expectation values, are those that satisfy the axiom of reflection positivity \cite{gj81}.  This is an extremely tight constraint.  However, it is a necessary but not sufficient condition.

The question of light-quark confinement can be translated into that of charting the infrared behavior of QCD's \emph{universal} $\beta$-function.  It is important to appreciate that while this function may depend on the scheme chosen to renormalize the quantum field theory, it is unique within a given scheme.  An elemental goal of hadron physics during the next ten years must be to design a program of experiment and theory that can together map out the $\beta$-function.  This is a well-posed problem.  It's importance is already widely appreciated and an exploratory attempt has been made \cite{Deur:2005cf}.

\begin{figure}[t]
%\vspace*{1cm}

\leftline{\includegraphics[clip,width=8cm]{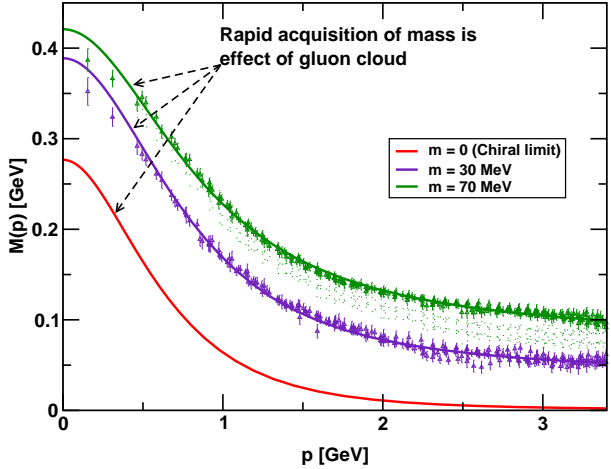}}
\vspace*{-7cm}

\rightline{
\parbox{8cm}{\caption{\label{gluoncloud} \emph{Dressed-quark mass function, 
      $M(p)$: solid curves -- DSE results \protect\cite{Bhagwat:2003vw,Bhagwat:2006tu}, 
     ``data'' -- numerical simulations of unquenched lattice-QCD 
      \protect\cite{Bowman:2005vx}.  In this figure one observes the current-quark 
      of perturbative QCD evolving into a constituent-quark as its momentum becomes 
      smaller.  The constituent-quark mass arises from a cloud of low-momentum gluons 
      attaching themselves to the current-quark.  This is dynamical chiral symmetry 
      breaking: an essentially nonperturbative effect that generates a quark mass 
      \emph{from nothing}; namely, it occurs even in the chiral limit.}}}
}

\end{figure}

While light-quark confinement remains a conjecture, many statements of 
fact can be made in connection with DCSB.  For example, DCSB explains the origin of constituent-quark masses and underlies the success of chiral effective field theory.  Understanding DCSB within QCD proceeds from the renormalised gap equation \cite{Maris:1997hd}:
\begin{equation}
S(p)^{-1} =  Z_2 \,(i\gamma\cdot p + m^{\rm bm}) + 
Z_1 \int^\Lambda_q\! g^2 D_{\mu\nu}(p-q) \frac{\lambda^a}{2}\gamma_\mu 
S(q) \Gamma^a_\nu(q,p)\frac{d^{4}q}{(2\pi)^{4}} , \label{gendse}
\end{equation}
where $\int^\Lambda_q$ represents a Poincar\'e invariant regularisation of the integral, with $\Lambda$ the regularisation mass-scale, $D_{\mu\nu}$ is the renormalised dressed-gluon propagator, $\Gamma_\nu$ is the renormalised dressed-quark-gluon vertex, and $m^{\rm bm}$ is the quark's $\Lambda$-dependent bare current-mass.  The vertex and quark wave-function renormalisation constants, $Z_{1,2}(\zeta^2,\Lambda^2)$, depend on the gauge parameter.  

The solution to Eq.\,(\ref{gendse}) has the form
\begin{eqnarray} 
%\nonumber 
 S(p) & =&  -i \gamma\cdot p \,\sigma_V(p^2,\zeta^2) + \sigma_S(p^2,\zeta^2) = \frac{1}{i \gamma\cdot p \, A(p^2,\zeta^2) + B(p^2,\zeta^2)} =
%= \frac{Z(p^2,\zeta^2)}{i\gamma\cdot p + M(p^2)} \\
\frac{Z(p^2,\zeta^2)}{i\gamma\cdot p + M(p^2)}\,
%
%& = &  - i \gamma\cdot p \,\sigma_V(p^2,\zeta^2) + \sigma_S(p^2,\zeta^2) \,.
\label{Sgeneral}
\end{eqnarray} 
and it is important that the mass function, $M(p^2)=B(p^2,\zeta^2)/A(p^2,\zeta^2)$ is independent of the renormalisation point, $\zeta$.  

The dressed-quark mass function in QCD is depicted in Fig.\,\ref{gluoncloud}.  It is one of the most remarkable features of the theory.  In perturbation theory it is impossible in the chiral limit to obtain $M(p^2)\neq 0$: the generation of mass \emph{from nothing} is an essentially nonperturbative phenomenon.  On the other hand, it is a longstanding prediction of nonperturbative DSE studies that DCSB will occur so long as the integrated infrared strength possessed by the gap equation's kernel exceeds some critical value \cite{Roberts:1994dr}.  There are strong indications that this condition is satisfied in QCD \cite{Bhagwat:2003vw,Bhagwat:2006tu,Bowman:2005vx}.  It follows that the quark-parton of QCD acquires a momentum-dependent mass function, which at infrared momenta is $\sim 100$-times larger than the current-quark mass.  This effect owes primarily to a dense cloud of gluons that clothes a low-momentum quark \cite{Bhagwat:2007vx}.  It means that the Higgs mechanism is largely irrelevant to the bulk of normal matter in the universe.  Instead the single most important mass generating mechanism for light-quark hadrons is the strong interaction effect of DCSB; e.g., one can identify it as being responsible for 98\% of a proton's mass. 

It is widely anticipated that there is an intimate connection between DCSB and light-quark confinement.  For example, analogous to quenched QCD, quenched QED in three dimensions (two spacial, one temporal -- QED$_3$) is confining because it has a nonzero string tension \cite{Gopfert:1981er}.  The effect of unquenching; viz., allowing light fermions to influence the theory's dynamics, has been much studied.  The nature of QED$_3$ is such that there is almost certainly a critical number of light flavors above which DCSB is impossible.  Moreover, chiral symmetry restoration and deconfinement are coincident owing to an abrupt change in the analytic properties of the fermion propagator when a nonzero scalar self-energy becomes insupportable \cite{Bashir:2008fk}.

The complex of Dyson-Schwinger equations (DSEs) is a powerful tool that has been employed with marked success to study confinement and DCSB, and their impact on hadron observables \cite{Roberts:1994dr,Roberts:2000aa,Alkofer:2000wg,Maris:2003vk,Roberts:2007jh}.  Moreover, the existence of a nonperturbative and symmetry preserving truncation scheme \cite{Munczek:1994zz,Bender:1996bb,Bender:2002as,Bhagwat:2004hn} has enabled the DSEs to be used to provide an explanation of dynamical chiral symmetry breaking and prove a body of exact results for pseudoscalar mesons \cite{Maris:1997hd,Maris:1997tm}.  They relate even to radial excitations and/or hybrids \cite{Holl:2004fr,Holl:2005vu,McNeile:2006qy}, and heavy-light \cite{Ivanov:1997yg,Ivanov:1998ms} and heavy-heavy mesons \cite{Bhagwat:2006xi}.  Mesons are described by the fully covariant Bethe-Salpeter equation and the exact results have been illustrated using a renormalisation-group-improved ladder-rainbow truncation of this and the gap equation \cite{Maris:1997tm,Maris:1999nt}, which also provided a prediction of the electromagnetic pion form factor \cite{Maris:2000sk}.  (Ladder-rainbow is the leading-order DSE truncation.)  In addition, algebraic parametrizations of the dressed-quark propagators and meson bound-state amplitudes obtained from such studies continue to be useful, in particular with the study of $B$-meson $\to$ light-meson transition form-factors \cite{Ivanov:2007cw} and baryon properties \cite{Alkofer:2004yf,Holl:2005zi,Flambaum:2005kc,Cloet:2008wg}.

In quantum field theory a baryon appears as a pole in a six-point quark Green function.  The residue is proportional to the baryon's Faddeev amplitude, which is obtained from a Poincar\'e covariant Faddeev equation that sums all possible exchanges and interactions that can take place between three dressed-quarks.  A tractable Faddeev equation for baryons was formulated in Ref.\,\cite{Cahill:1988dx}.  It is founded on the observation that an interaction which describes colour-singlet mesons also generates quark-quark (diquark) correlations in the colour-$\bar 3$ (antitriplet) channel \cite{Cahill:1987qr}.  The lightest diquark correlations appear in the $J^P=0^+,1^+$ channels and hence only they are retained in approximating the quark-quark scattering matrix.  While diquarks do not appear in the strong interaction spectrum; e.g., Refs.\,\cite{Bender:1996bb,Bender:2002as,Bhagwat:2004hn}, the attraction between quarks in this channel justifies a picture of baryons in which two quarks are always correlated as a colour-$\bar 3$ diquark pseudoparticle, and binding is effected by the iterated exchange of roles between the bystander and diquark-participant quarks.

\begin{figure}[t]

\leftline{\includegraphics[clip,width=22em]{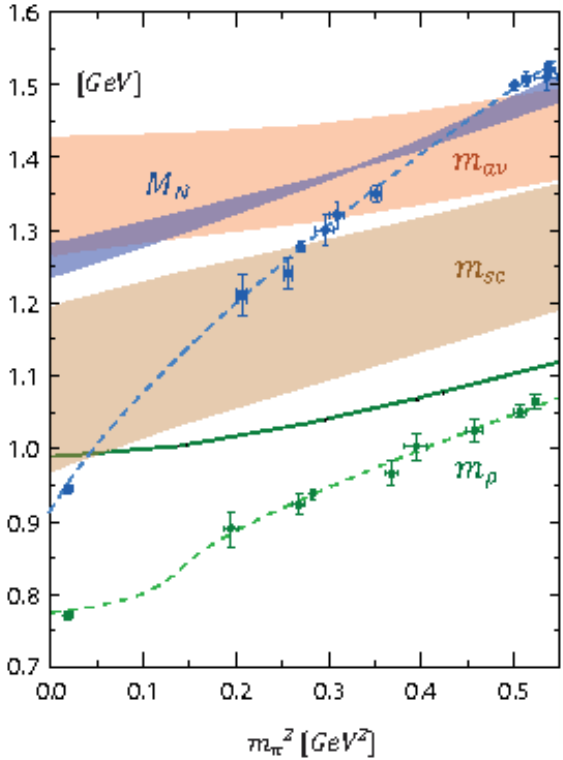}}
%\leftline{\parbox{8cm}{\Large MISSING FIGURE
%          \\ . \\ . \\ . \\ . \\ . \\ . \\ . \\ . \\ . \\ . \\ . 
%          (corruped file mNomega.eps)}}
\vspace*{-11cm}
\rightline{\parbox{8cm}{\caption{\label{nucleonmass} 
\emph{Thick bands}: Evolution with current-quark mass, $\hat m$, of the scalar and axial-vector diquark masses: $m_{sc}$ and $m_{av}$.  Bands demarcate sensitivity to the variation in $\omega$: $r_a=1/\omega$ can be associated with a confinement length-scale in the quark-quark scattering kernel.  ($m_\pi$, calculated from rainbow-ladder meson Bethe-Salpeter equation: $\hat m = 6.1\,$MeV $\Rightarrow m_\pi = 0.138\,$GeV.)
\emph{Solid curve}: Evolution of $\rho$-meson mass \protect\cite{Eichmann:2008ae}.  This observable quantity is insensitive to $\omega$.  
With $m_\rho$, results from simulations of lattice-regularised QCD \protect\cite{AliKhan:2001tx} are also depicted along with an analysis and chiral extrapolation \protect\cite{Allton:2005fb}, \emph{short dashed curve}.  
\emph{Thin band}: Evolution with $\hat m$ of the nucleon mass obtained from the Faddeev equation: $\hat m =6.1\,$MeV, $M_N=1.26(2)\,$GeV cf.\ results from lattice-QCD \cite{Ali Khan:2003cu,Frigori:2007wa} and an analysis of such results \cite{Leinweber:2003dg}, \emph{dashed curve}.  (Figure adapted from Ref.\,\cite{Eichmann:2008ef}.)
}}
} 
\end{figure}

The Poincar\'e covariant and quantum field theoretical DSE framework is well suited to addressing the question of light-quark confinement.  It may be posed as the problem of developing a detailed understanding of the infrared evolution of the quark-quark scattering kernel, $K_{q \bar q}$.  With Refs.\,\cite{Eichmann:2008ae,Eichmann:2008ef} significant progress has been made in this direction.  They enable the direct correlation of meson and baryon properties via a single interaction kernel that preserves QCD's one-loop renormalisation group behaviour and can systematically be improved.  The unified framework provides a veracious description of the pion as both a Goldstone mode and a bound state of dressed-quarks.  It is the only approach that is capable of doing so because it alone is capable of expressing the behavior in Fig.\,\ref{gluoncloud}.  The studies predict, amongst other things, the evolution of the nucleon mass with a quantity that can methodically be connected with the current-quark mass in QCD.  This is depicted in Fig.\,\ref{nucleonmass}.   Notably, the nucleon mass is insensitive to the kernel's single parameter despite the large dependence of the unobservable diquark masses.  Systematic corrections to the DSE's leading order truncation have been shown to move results into line with experiment.

An international theory program is underway that exploits the strengths of the DSEs in studies of the spectrum and interactions of hadrons.  In connection with this, a comprehensive study of nucleon electromagnetic form factors has just been completed \cite{icloetNFF}.  It evaluates a dressed-quark core contribution, which is defined by the solution of a Poincar\'e covariant Faddeev equation in which dressed-quarks provide the elementary degree of freedom and correlations between them are expressed via diquarks.  The diquarks are nonpointlike and the current depends on their charge radii.  A particular feature of the study is a separation of form factor contributions into those from different diagram types and correlation sectors, and subsequently a flavour separation for each of these.  Amongst the extensive body of results that one might highlight: $r_1^{n,u}>r_1^{n,d}$, owing to the presence of axial-vector quark-quark correlations; and for both the neutron and proton the ratio of Sachs electric and magnetic form factors possesses a zero.

\begin{figure}[t]
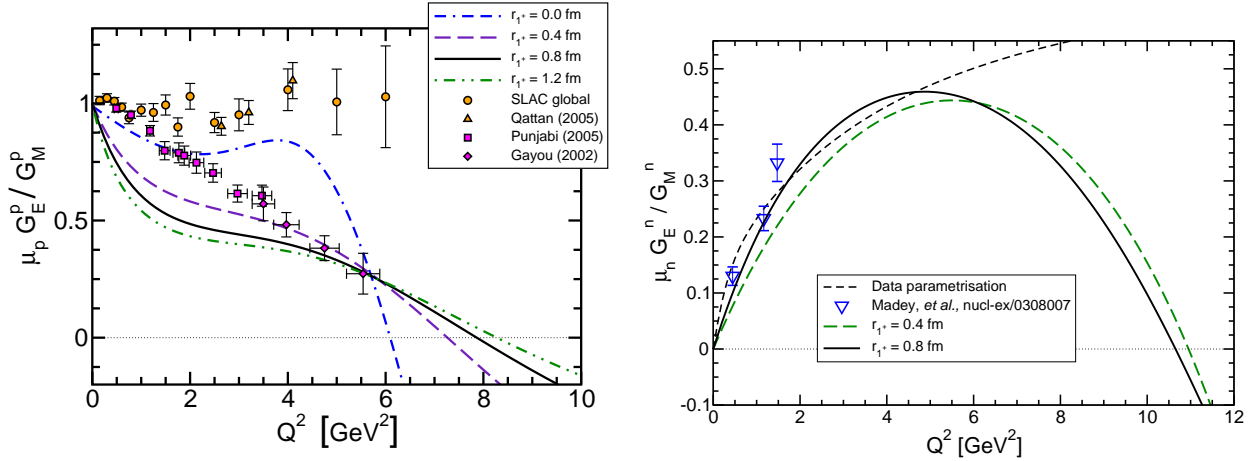

\leftline{\includegraphics[width=8cm]{DSE/FIGS/GEGM0809.eps}}
\vspace*{-5.5cm}

\rightline{\includegraphics[width=8cm]{DSE/FIGS/GEGMn.eps}}
\caption{\label{fig:GEGM} \emph{Left panel} --  Result for the normalised ratio of proton Sachs electric and magnetic form factors computed with four different diquark radii.  
Data: diamonds -- \cite{gayou}; squares -- \cite{Punjabi:2005wq}; triangles -- \cite{Qattan:2004ht}; and circles \cite{Walker:1993vj}.  \emph{Right panel} -- Analogous ratio for the neutron computed with two different diquark radii.  
\emph{Short-dashed curve}: parametrisation of Ref.\,\protect\cite{Kelly:2004hm}.  \emph{Down triangles}: data from Ref.\,\protect\cite{Madey:2003av}.
} 
\end{figure}

The latter ratios are depicted in Fig.\,\ref{fig:GEGM}.  A sensitivity to the nucleon's electromagnetic current is evident, here expressed via the diquarks' radius.  However, irrespective of that radius, the electric form factors possess a zero and the magnetic form factor is positive definite.  On $Q^2\lsim 3\,$GeV$^2$ the proton result lies below experiment.  As explained in Ref.\,\cite{icloetNFF}, this can likely be attributed to omission of so-called pseudoscalar-meson-cloud contributions.  

\begin{figure}[t]
\leftline{\includegraphics[width=8cm]{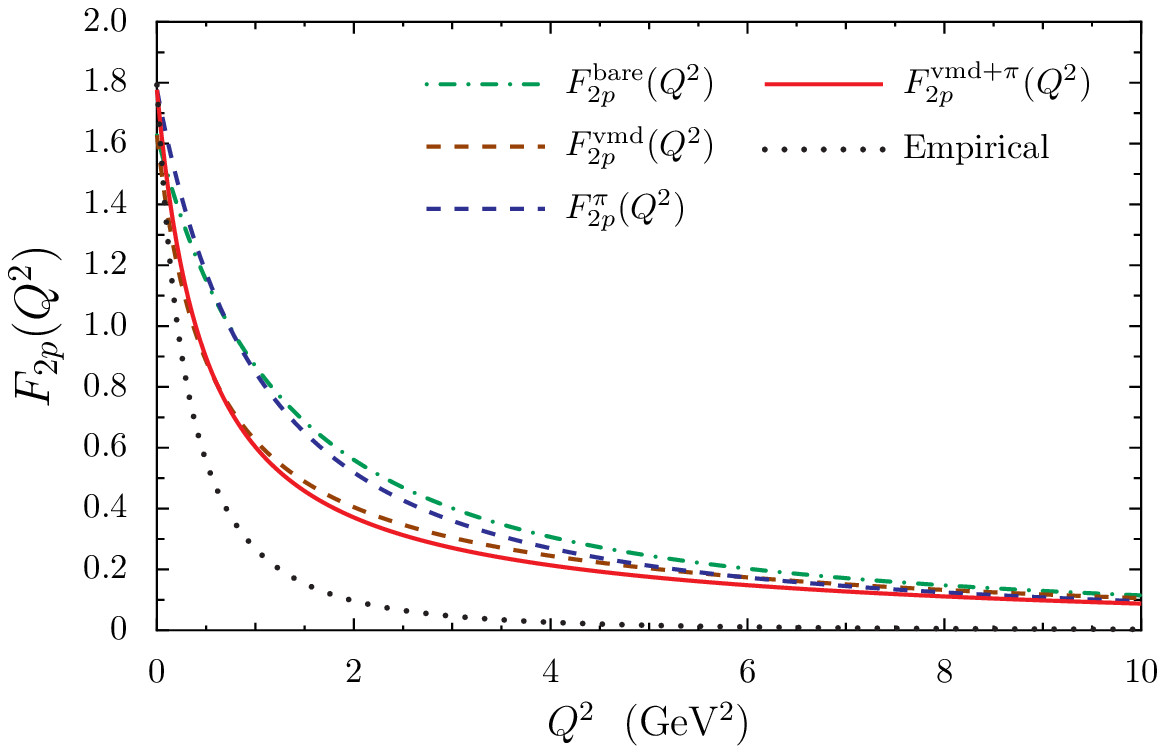}}
\vspace*{-5.5cm}

\rightline{\includegraphics[width=8cm]{DSE/FIGS/F2pnDSEKelly}}
\caption{\label{NJLcfDSE} \emph{Left panel} --  Confining-NJL model Faddeev equation result for the proton's Pauli form factor: \emph{solid curve}, complete result; and \emph{dotted curve}, parametrization of experimental data \cite{Kelly:2004hm}.  The curves labelled \emph{bare}, \emph{VMD} and $\pi$ represent intermediates stages in the calculation of the solid curve.  \emph{Right panel} -- Difference between a DSE-calculated dressed-quark core contribution to the Pauli form factor and a parametrisation of experimental data \protect\cite{Kelly:2004hm}, each normalised by the appropriate anomalous magnetic moment at $Q^2=0$: \emph{dashed curve} -- proton; \emph{solid curve} -- neutron.  At $Q^2\approx 2 M_N^2$ the difference between calculation and data in the \emph{left panel} is an order of magnitude larger than in the \emph{right panel}.
} 
\end{figure}

It has long been recognized that the behavior characterized by Fig.\,\ref{gluoncloud} has an enormous impact on hadron phenomena \cite{Roberts:1994hh} and hence that a form factor's pointwise evolution with momentum transfer is a sensitive probe of the nature of the quark-quark scattering kernel.  For example, this was made strikingly apparent for the pion in Ref.\,\cite{Maris:1998hc}.  It can also be seen for the nucleon.  
In the left panel of Fig.\,\ref{NJLcfDSE} we depict the proton's Pauli form factor calculated in a confining Nambu--Jona-Lasinio model, whose simplicity and phenomenological efficacy has recently been much exploited \cite{Cloet:2005rt,Cloet:2005pp,Cloet:2006bq,Cloet:2007em}.  This model possesses a dressed-quark mass but it \emph{does not run}; i.e., it assumes a large value that is momentum independent.  As apparent in the figure, in this case the agreement between model result and experiment deteriorates quickly with increasing momentum transfer and the ultraviolet power-law behavior is incorrect.  This may be contrasted with the behavior in the right panel, which is obtained \cite{icloetNFF} using a momentum-dependent running quark mass of the type depicted in Fig.\,\ref{gluoncloud}.  This calculation omits the pseudoscalar meson cloud.  However, it retains the fully momentum dependent dressed-quark structure, which ensures good agreement with data for $Q^2 \approx 2\,$--$\,3 M_N^2$.

We judge that it is possible to employ precision data on nucleon-resonance transition form factors as a means by which to chart the momentum evolution of the dressed-quark mass function and therefrom the infrared behavior of QCD's $\beta$-function; in particular, to locate unambiguously the transition boundary between the constituent- and current-quark domains that is signaled by the sharp drop apparent in Fig.\,\ref{gluoncloud}.  That can be related to an inflexion point in QCD's $\beta$-function.  Contemporary theory indicates that this transition boundary lies at $p^2 \sim 0.6\,$GeV$^2$.  Since a probe's input momentum $Q$ is principally shared equally amongst the dressed-quarks in a transition process, then each can be considered as absorbing a momentum fraction $Q/3$.  Thus in order to cover the domain $p^2\in [0.5,1.0]\,$GeV$^2$ one requires $Q^2\in [5,10]\,$GeV$^2$.

An international theory effort is underway in order to realize the goal of turning experiment into a probe of the dressed-quark mass function.  The effort has many facets and the first calculations are being performed at leading-order in the DSE truncation.  

\begin{figure}[t]
\leftline{\includegraphics[width=8cm]{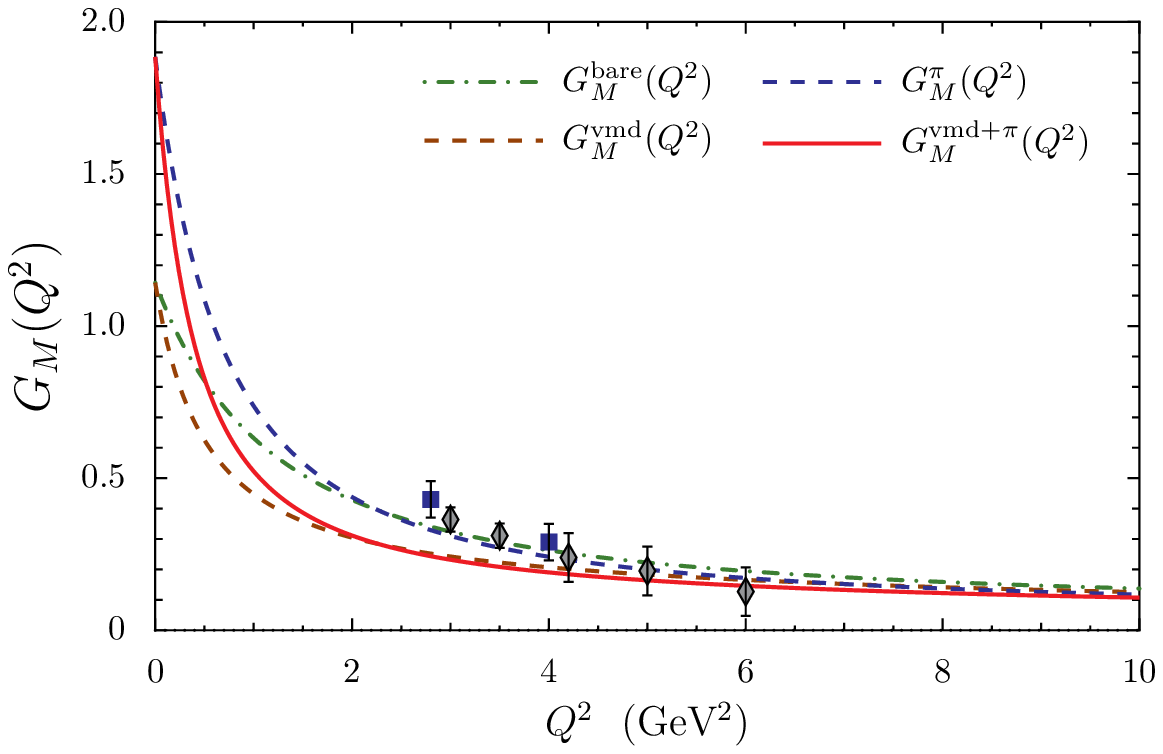}}
\vspace*{-4.5cm}
\rightline{
\parbox{8cm}{\caption{\label{ICNDelta} \emph{Solid curve} -- Confining-NJL model Faddeev equation result for the $N\to \Delta$ M1 transition form factor, complete calculation. The curves labelled \emph{bare}, \emph{VMD} and $\pi$ represent intermediates stages in the calculation.  Data from Refs.\,\protect\cite{Frolov:1998pw,Ungaro:2006df}.
}}
}
\vspace*{1cm}
 
\end{figure}

Naturally, a reference calculation is needed, one that does not incorporate the running of the dressed-quark mass which is such a singular feature of QCD.  A calculation of this type is nearing completion \cite{icloetNDelta} and a preliminary result is presented in Fig.\,\ref{ICNDelta}.  It is evident that the pion is playing a very important role but significant strength is missing in the neighborhood of $Q^2=0$, since empirically $G_M(Q^2=0)=3$.  This calculation must be analyzed and the origin of each feature and defect determined so that the role of a constant constituent-quark-like mass can unambiguously be identified.  The analysis should be complete by mid-2009.

Following this effort the Faddeev equation framework of Refs.\,\cite{Alkofer:2004yf,Holl:2005zi,Flambaum:2005kc,Cloet:2008wg,icloetNFF}, described briefly above and widely employed in studies of nucleon and $\Delta$ properties, will be applied to the $N\to\Delta$ transition.  The strong momentum dependence of the dressed-quark mass function is an integral part of this framework.  Therefore, in this study it will be possible, e.g., to vary artificially the position of the marked drop in the dressed-quark mass function and thereby identify experimental signatures for its presence and location. 
 This study would begin in 2010 and be completed by the end of that year.

In parallel with these efforts, the \emph{ab-initio} 
rainbow-ladder DSE framework of Refs.\,\cite{Eichmann:2008ae,Eichmann:2008ef} 
is being extended to the $\Delta$ resonance.  A solution of the Faddeev 
equation for the $\Delta$ should be complete by the end of 2010 \cite{DNprivate}.  
The nucleon-photon current developed in Ref.\,\cite{Eichmann:2008ef} 
will then be generalized so that its nucleon form factor studies can be 
correlated with a calculation of the $N\to\Delta$ transition.  
The time required to complete this effort is uncertain, given that it involves 
a PhD student who is now nearing completion of his research, 
but assuming that a new student is found or a postdoctoral fellow can assume 
responsibility, a reasonable estimate is for completion by the end of 2010.  
It should be emphasized, however, that for technical reasons this effort can 
only produce form factors out to modest 
momentum transfer; viz., $Q^2\sim 2\,M_N^2$.  

In order to extend the calculations it is imperative to improve the numerical methods used in the calculation of form factors and also to improve the rainbow-ladder quark-quark scattering kernel.  This is naturally part of the next phase of the theoretical effort.    

One should also proceed beyond the leading-order DSE truncation.  This is necessary in order to identify and isolate artefacts that may arise through truncation and their impact on predictions for experimental signatures of the transition between the constituent-quark and the current-quark domains.  This need notwithstanding, the merits of the rainbow-ladder truncation should not be underestimated.  It is exact for $p^2\gsim 1\,$GeV$^2$.  Furthermore, contemporary estimates show that at smaller $p^2$ it is still semi-quantitatively accurate for a wide range of observables, the nature of which can be determined \emph{a priori}.  Careful application of the rainbow-ladder truncation yields insights that are generally reliable.

A path for proceeding beyond the rainbow-ladder truncation is charted.  Owing to the relative ease of dealing with the Bethe-Salpeter equation, it will initially proceed via mesons.  The one-parameter model for the infrared behavior of $K_{q \bar q}$ in Ref.\,\cite{Eichmann:2008ae} will be employed in in DSE calculations of the spectrum and interactions of pseudoscalar mesons with masses $<2\,$GeV.  Comparison with scant extant data will inform improvements of the \emph{Ansatz}, as will continuing DSE and lattice-QCD research on the pointwise behavior of the dressed-quark-gluon vertex.  It is in a nontrivial vertex that one moves beyond the rainbow-ladder truncation.  

The improved $K_{q \bar q}$ will be employed in studies of the spectrum and interactions of axial-vector mesons, all of which lie above $1\,$GeV.  The properties of pseudoscalar excited states and axial-vector mesons are a sensitive probe of the long-range part of the interaction between light-quarks.  Comparison with scarce data will assist in further improving the map of the light-quark confinement interaction.  A well constrained form of $K_{q \bar q}$ will thereafter be available.  It will enable reliable predictions for the properties of all mesons in the $1-2\,$GeV range, including hybrids and exotics.  This extended kernel will provide the basis for future \emph{ab initio} Faddeev equation studies of the nucleon and $\Delta$.  One may anticipate that those studies could begin in 2013.

In the meantime, following the successful completion of $N\to \Delta$ studies, the dressed-quark Faddeev equation will be employed in nucleon resonance spectroscopy and the calculation of additional nucleon to resonance transitions.  The starting point for this effort will be a calculation of the dressed-quark component of the Roper resonance.  With experiment \cite{cqm-rel-4} now pointing to an interpretation of the $N(1440)$ as a radial excitation of the nucleon, a compelling case can be made for employing a quantum field theoretical approach to QCD that is founded on dressed-quark degrees of freedom in order to determine whether the experimental claim is consistent with the best available theory.  A conclusion on this point should be available from the DSE-based Faddeev equation by the end of 2011 and from the \emph{ab initio} rainbow-ladder truncation by 2012.  

In parallel with the program outlined here an effort will be underway at the Excited Baryon Analysis Center (EBAC), which will provide the reaction theory necessary to make reliable contact between experiment and predictions based on the dressed-quark core.  While rudimentary estimates can and will be made of the contribution from pseudoscalar meson loops to the dressed-quark core of the nucleon and its excited states, a detailed comparison with experiment will only follow when the DSE-based results are used to constrain the input for dynamical coupled channels calculations.

%\end{document}

%  \newpage
\section{ Electroproduction of $N^*$ resonances at large momentum transfers
 }
  % contribution to the White Paper of October NSTAR JLAB Workshop

%\documentclass[aps,prl,twocolumn,groupedaddress,superscriptaddress]{revtex4}
%\documentclass[aps,prd]{revtex4}

%\usepackage[utf8]{inputenc}
%\usepackage[T1]{fontenc}
%\usepackage{slashed}
%\usepackage{times}

%\usepackage{amssymb,amsfonts,amsmath,amsthm}

%\usepackage{slashed}
%\usepackage{dcolumn}
%\newcolumntype{l}[1]{D{.}{\cdot}{#1}} 

%\usepackage{graphicx}
%\usepackage[tight,hang,raggedright,normalsize]{subfigure}

%%%%%%%%%%%%%%%%%%%%%%%%%

\newcommand{\dev}{$\chi^2/\textrm{d.o.f}$}
\newcommand{\msb}{$\overline{\text{MS}}$}

\newcommand{\nn}{\nonumber}
\newcommand{\up}{{\uparrow}}
\newcommand{\down}{{\downarrow}}

%%%%%%%%%%%%%%%%%%%%%%%%%

%\begin{document}

%\begin{center}
%%{\bf Electroproduction of $N^*$ resonances\\ at large momentum transfers}
%%\\
%%\vspace{0.5cm}
%{\large Vladimir Braun}
%%%{\it Institut f\"ur Theoretische Physik, Universit\"at Regensburg, 93040 Regensburg, Germany}
%\end{center}

%%%%%%%%%%%%%%%%%%%%%%%%%%%%%%%%

Form factors play an extremely important role in the
studies of the internal structure of composite particles as the measure
of charge and current distributions.
In particular transitions  to nucleon excited states 
allow to study the relevant degrees of freedom, 
wave function and interaction between the constituents, 
and the transition to pQCD.
 The prediction of QCD is that at large momentum transfers 
the form factors become increasingly dominated by the 
contribution of the valence state 
with small transverse separation between the quarks. 
There is a growing consensus that the ultimate pQCD picture based on hard 
rescattering involving two gluon exchanges 
%purely perturbative hard factorization regime 
is not achieved at present energies; however, 
at photon virtualities from 5 to 10 GeV$^2$ of CLAS12 we 
will have access to quark degrees of freedom, whereas the 
description in terms of meson-baryon degrees of freedom becomes
much less suitable than at smaller momentum transfers. 
%
%the emergence of QCD current quarks and gluons 
%as the adequate degrees of freedom is expected to happen earlier, 
%at $Q^2\sim$ a few GeV$^2$. 
%In this transition region the form factors  
%can be measured to high accuracy, and such measurements can be part of the 
%experimental programm for the 12 GeV upgrade at Jefferson Lab.

The major challenge for theory is that quantitative description of 
form factors in this region must include soft nonperturbative 
contributions. An approach that is most directly connected to QCD
is based on the light-cone sum rules (LCSRs)~\cite{Braun:2001mu,Braun:2006hz}.
This technique allows one to calculate form factors using much more limited 
information compared  to the full nonperturbative wave functions, albeit 
with some assumptions. 
The LCSRs are derived from the correlation function of the type
\begin{equation}
\int\! dx\, e^{-iqx}\langle N^*(P)| T \{ \eta (0) j_\mu^{\rm em} (x) \} | 0 \rangle
\end{equation}
where $\eta$ is a suitable operator with nucleon quantum numbers. More detail can be found in the following contribution to these proceedings (Ref.~\cite{stoler-WP}). 
Making  use of the duality of QCD quark-gluon and hadronic degrees of freedom  
through dispersion relations one can write a representation for the 
transition form factors in terms of the  $N^*$
momentum fraction distributions of partons at small transverse separations
in the $N^*$, dubbed distribution amplitudes (DAs) 
which are the same quantities 
that enter the pQCD calculation, cf.~\cite{Carlson:1985mm,Carlson:1988gt}. 
The LCSRs provide one with the most direct relation of the hadron 
form factors and DAs that is available at present, with no other 
nonperturbative parameters.  

\begin{figure}[ht]

\subfigure[\label{fig:nlat}]{\includegraphics[width=0.47\textwidth,clip]{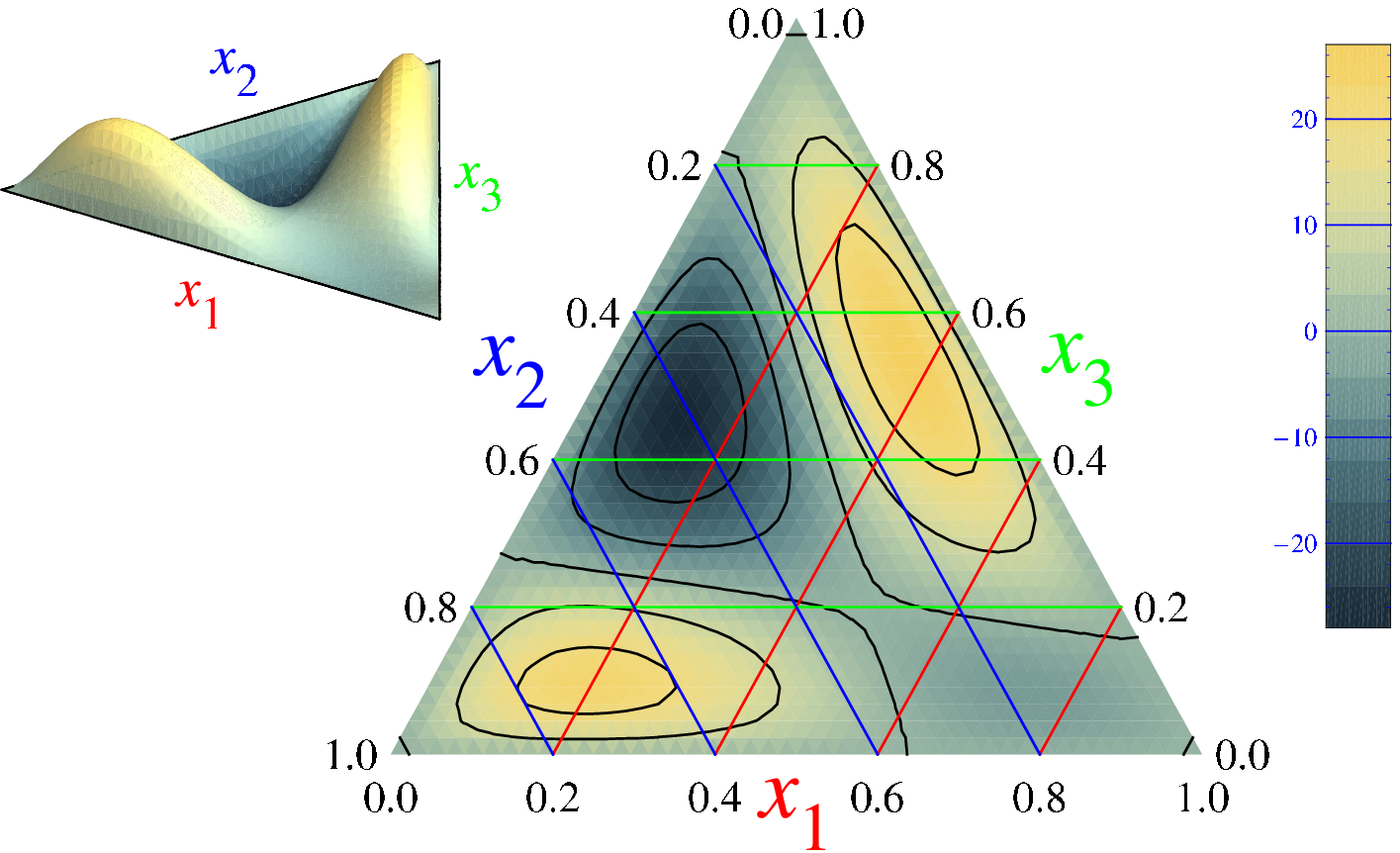}}

\subfigure[\label{fig:nstar}]{\includegraphics[width=0.47\textwidth,clip]{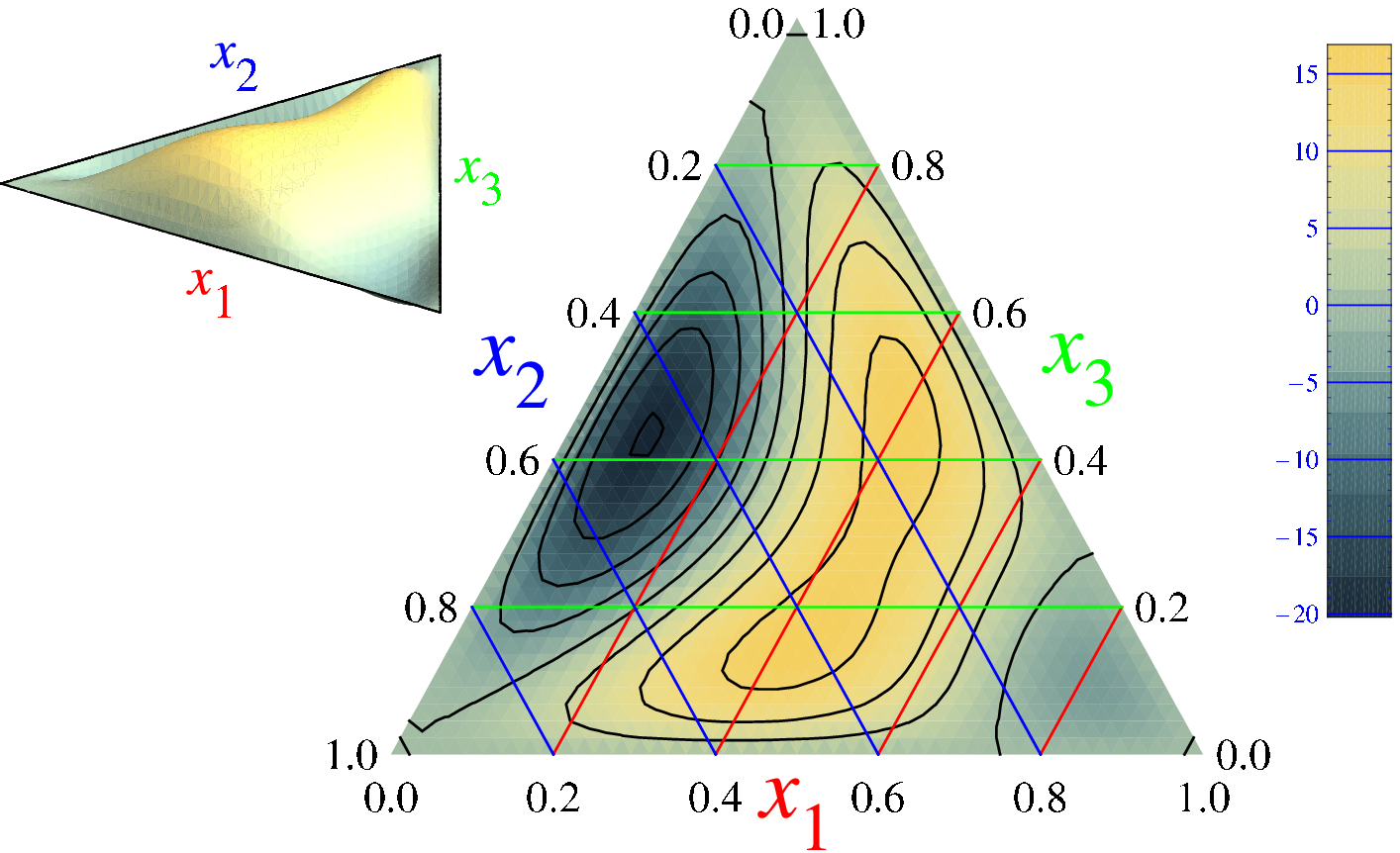}}
\caption{
Barycentric plot of the distribution amplitudes for nucleon (a) and $N^\star(1535)$ 
(b) at $\mu_{\overline{MS}}=1\mathrm{GeV}$~\cite{Warkentin:2008iu}.
The lines of constant $x_1$, $x_2$ and $x_3$ are parallel 
to the sides of the triangle labelled by $x_2$, $x_3$ and $x_1$, respectively.
}
\label{fig:barycentric}
\end{figure}

The necessary information on the DAs can be obtained from LQCD.
The theoretical particle and nuclear physics group in Regensburg is a member 
of QCDSF and the SFB/Transregio 55 ``Hadron Physics with Lattice QCD'' which 
is a large-scale research program aimed at the study of hadron structure using 
LQCD techniques. The studies of hadron DAs present 
one of the long-term goals of this collaboration and they will be continued. 
The most important steps so far have been the calculation of the second moment
of the pion DA \cite{Braun:2006dg}, the classification of three-quark operators 
in irreducible spinor respresentations of the hypercubic group
\cite{Kaltenbrunner:2008pb}, the calculation of nonperturbative renormalization 
constants for three-quark operators \cite{Gockeler:2008we} and 
the evaluation of the lowest moments of the DAs of the nucleon~\cite{Braun:2008ur}
and its parity partner $N^*(1535)$~\cite{Warkentin:2008iu}.
According to our preliminary study, quark distributions in the nucleon 
and $N^*(1535)$ are rather different, see Fig.~\ref{fig:barycentric}. 

We find a larger wave function of the three quarks at the origin in the $J^P=\frac12^-$ state 
compared to $J^P=\frac12^+$ state that may be counterintuitive.
The momentum fraction carried by the $u$-quark with the same helicity as the baryon 
itself appears to be considerably larger for $N^*$ resonance, 
indicating that its DA is more asymmetric.
The future plans are, first of all, to repeat the same calculations 
with smaller pion masses and larger lattices that are expected to become
available within 2-3 years. This would remove a major source of uncertainties
which is due to the chiral extrapolation. Second, we want to expand our calculation
of the moments of the DAs to the whole $J^P=1/2^+$ and $J^P=1/2^-$ baryon octets and later 
also to the decuplet. We will also explore possibilities to calculate higher moments
of DAs and also moments of the generalized parton distributions involving different      
hadrons in the initial and final state (sometimes referred to as TDAs).
All such calculations will require a dedicated effort, and the accuracy of the 
predictions may vary.

\begin{figure}[ht]

\includegraphics[width=0.55\textwidth,clip]{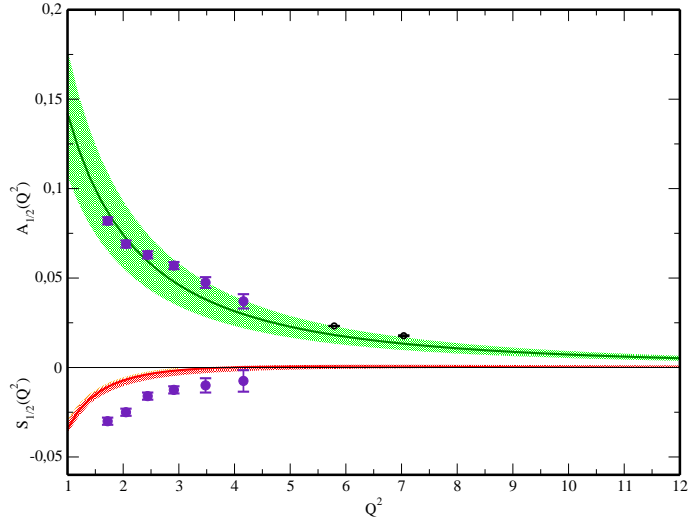}
\caption{The LCSR calculation for the helicity amplitudes
$A_{1/2}(Q^2)$ and $S_{1/2}(Q^2)$ for the electroproduction of $N^*(1535)$ resonance 
using lattice results for the $N^*$ distribution amplitudes~\cite{nstar} compared to
the available experimental data. The points at $Q^{2}$ $<$ 5.0 GeV$^2$ are from
the CLAS data analysis~\cite{Bu08c}. 
The points at $Q^{2}$ $>$ 5.0 GeV$^2$  are the Hall C data~\cite{Dalton}, obtained 
under the assumption $A_{1/2}>>S_{1/2}$.
The curves are obtained using central values of lattice parameters and the 
shaded areas show the corresponding uncertainty.
}
\label{S11res}
\end{figure}

The results of the LCSR calculation of the helicity amplitides
of the electroproduction of $N^*(1535)$ using LCQD input on the DA are  
presented in Fig.~\ref{S11res}. 
This calculation  corresponds to the simplest, tree-level (or leading-order) LCSRs
The errors on the parameters of the DAs induce an uncertainty in the  
calculation of the form factors of order 30\%. This can be reduced in the future.
In addition, using quark-hadron duality in the identification of the nucleon
contribution, which is endemic to the LCSR approach, introduces an irreducible 
uncertainty of the order of 10-20\% in the whole $Q^2$ range.
In the region  $Q^2 > 2$~GeV$^2$ where the
light-cone expansion may be expected to converge, the results appear to
describe the general features of the data rather well. 
The small $S_{1/2}$ amplitude arises as a result of strong cancellations between
contributions of the helicity conserving and helicity violating form factors and
is difficult to predict reliably. The shown uncertainty 
is likely to be underestimated for this case.
%The disagreement for $A_{1/2}$ at momentum transfers below 2 GeV$^2$ is most
%likely due to large ``kinematic'' corrections involving powers of $m^2_{N^*}/Q^2$;
%this question will be addressed in future studies. 

The LCSR approach is rather general; it has 
been applied e.g. to the $N\gamma\Delta$ transitions \cite{Braun:2005be1} 
and to threshold pion electroproduction \cite{Braun:2006td1,Braun:2007pz1}.
In all cases, however, the LCSRs of the type considered here cannot be extended 
below $Q^2 \sim 1$~GeV$^2$ because of the so-called 
bilocal contributions to the operator product expansion.   

In order to match the expected accuracy of the next generation of lattice results, 
the LCSR calculations of baryon form 
factors will have to be advanced to include NLO radiative corrections, as it has 
become standard for meson decays. For the first effort in this direction, see
\cite{PassekKumericki:2008sj}.
In addition, it is necesary to develop a technique for the 
resummation of ``kinematic'' corrections to the sum rules 
that are due to nonvanishing masses of the resonances. The corresponding 
corrections to the total cross section of the deep-inelastic scattering
are known as Wandzura-Wilczek corrections and can be resummed to all 
orders in terms of the Nachtmann variable; we will be looking for a generalization
of this method to non-forward kinematics which is also important in
a broader context. 
With these improvements, we expect that the LCSR approach can be used to 
constrain light-cone DAs of the nucleon and its resonances from the 
comparison with the electroproduction data. These contraints can 
then be compared with the LQCD calculations.

%%%%%%%%%%%%%%%%%%%%%%%%%%%%%%%%

%\end{document}

%\newpage
\section{GPD and LCSR Representations of Resonance Form Factors } 
%\documentclass[11pt]{article}
% \usepackage{epsfig}
% \usepackage{times}
% \addtolength{\oddsidemargin}{-.875in}
%\addtolength{\evensidemargin}{-.875in}
%\addtolength{\textwidth}{1.75in}
%\addtolength{\topmargin}{-.875in}
%\addtolength{\textheight}{1.8in}

%\newcommand{\up}{{\uparrow}}
%\newcommand{\down}{{\downarrow}}

%\begin{document}

%\begin{centering}
%%\centerline{\bf \Large GPD and LCSR Representations of Resonance Form Factors} 

%%\vspace{0.1in}
%{\large Paul Stoler}

%%{\it Phyics Department, Rensselaer Polytechnic Institute, Troy, NY 12180}

%\end{centering}

%\vspace{0.1in}

 One of the primary goals of the JLab upgrade, and CLAS12 in particular, is to characterize the wave functions of the nucleon and its excitation in terms of the current quark and gluon fields. In principle, these wave functions can be constrained experimentally through measurements of exclusive reactions over large ranges of $x$ and $t$. Baryon elastic and transition form factors can be written as overlap integrals of the light-cone wave functions, and make an important contribution to this program. There are several approaches to encoding these overlap integrals in terms of the partonic degrees of freedom, i.e. $x$ and $t$, which connect them to the experimental data. Two examples we discuss here are  generalized parton distributions (GPD) and the light cone sum rule (LCSR), which were discussed in the previous contribution to these proceedings~\cite{Braun-WP}. In particular, we focus on how they specifically relate to the measured  form factors. These overlap integrals are the connecting points between theory and experiment. At this time the theoretical approach which most directly links QCD to these observables appears to be lattice QCD (LQCD). The goals of LQCD are to calculate the GPDs or DAs which can be fed into the basic relationships which predict the experimental results. 
  
\vspace{0.1in}

\noindent {\bf  GPDs and Resonance Form Factors.}

\vspace{0.1in}
The extraction of GPDs from experiments on  exclusive reactions at high momentum transfer, such as deeply virtual Compton scattering (DVCS) and deeply virtual meson production,  is one of the primary goals of the CLAS12 upgrade. Since elastic and baryon transition form factors  are the first moments of the GPDs,  they  provide important constraints  and thus provide a vital contribution to the overall exclusive reaction program.
The relationship of models of GPDs and elastic form factors have been  treated in detail, for example in Ref.~\cite{diehl-ff}. 

\vspace{0.1in}
 \noindent{\bf \large The $\bf  N \to \Delta(1232)$:}
 
The relationships between GPDs and resonance
form factors was worked out and treated several years ago in Refs. ~\cite{frankfurt}~\cite{goeke}.
The current structure of the transition  
$$\Gamma _{\nu \mu }  = G_M^* (q^2 ){K}_{\nu \mu }^M (q^2 ) + G_E^* (q^2 ){K}_{\nu \mu }^E (q^2 ) + G_C^* (q^2 ){K}_{\nu \mu }^C (q^2 )$$

\noindent leads to the following GPD relation:

\begin{eqnarray}\label{GPD-Delta}
\nonumber
\frac{P^+}{{2\pi }}\left. {\int {dy^ -  } e^{ix\bar P^ +  y^ -  } \left\langle {\Delta (p')} \right|\bar \psi_\Delta \left( { - y/2} \right)\gamma _\nu  n^\nu  \tau _3 \psi \left( {y/2} \right)\left| {N(p)} \right\rangle } \right|_{y^ +   = \vec y_ \bot   = 0}   
&=&\nonumber\\
\bar u^\beta_\Delta(p')\left\{ {H_M (q^2 ) {K}_{\beta \mu }^M (q^2 ) + H_E (q^2 ) {K}_{\beta\mu }^E (q^2 ) + H_C (q^2 ) {K}_{\beta \mu }^C (q^2 )}  \right\}  n^\mu u_p(p) 
\end{eqnarray}
 
\noindent In eq.  (\ref{GPD-Delta}) above, $u^\beta_\Delta(p')$ 
is a Rarita-Schwinger spinor for the $\Delta$, ${K}_{\beta \mu }^{M,E,C}$ are 
the covariants defined in~\cite{J-S},  $n^\mu$  is a light-cone vector 
normalized such that $n^2=0$ and $n^\mu P_\mu =1$. The relationship between the form factors and the GPDs is then

$$2G^* _M (t) = \int {dxH_M (t,x,\xi )},\ \ \ \ 2G^* _E(t) = \int {dxH_E(t,x,\xi )}\  \ {\rm and} \ \ \ 2G^* _C(t) = \int{dxH_C(t,x,\xi )}.$$

\begin{figure}[h]
\includegraphics[width=3.5in]{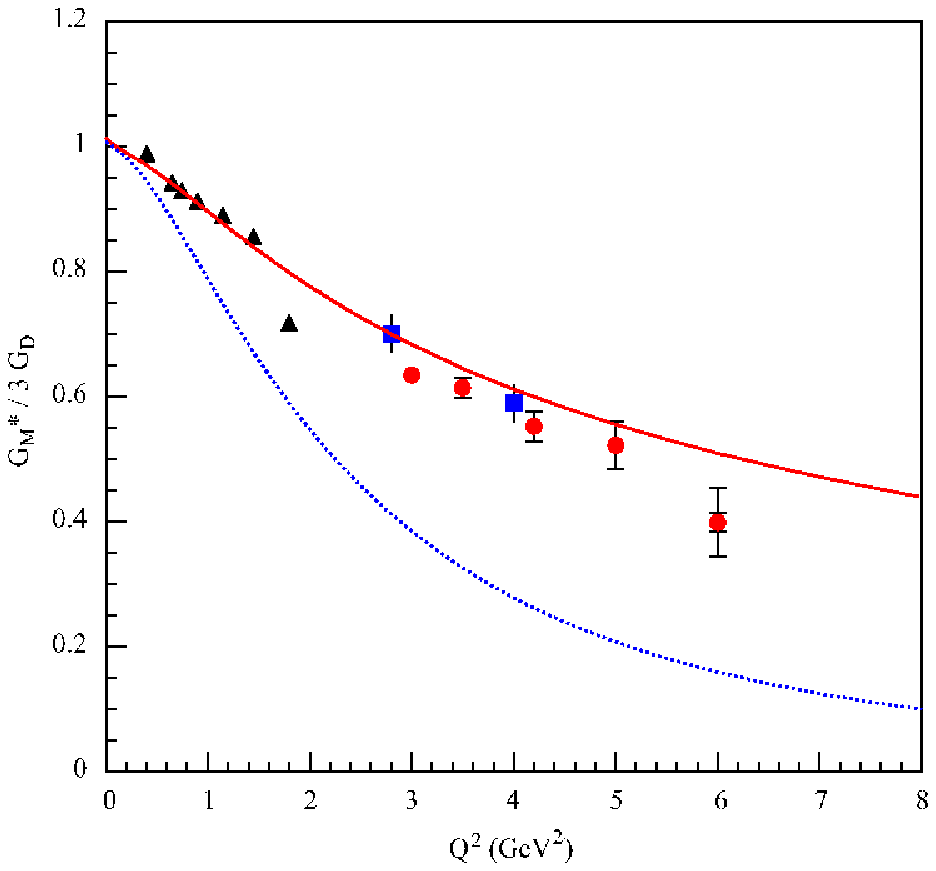}
\includegraphics[width=3.5in]{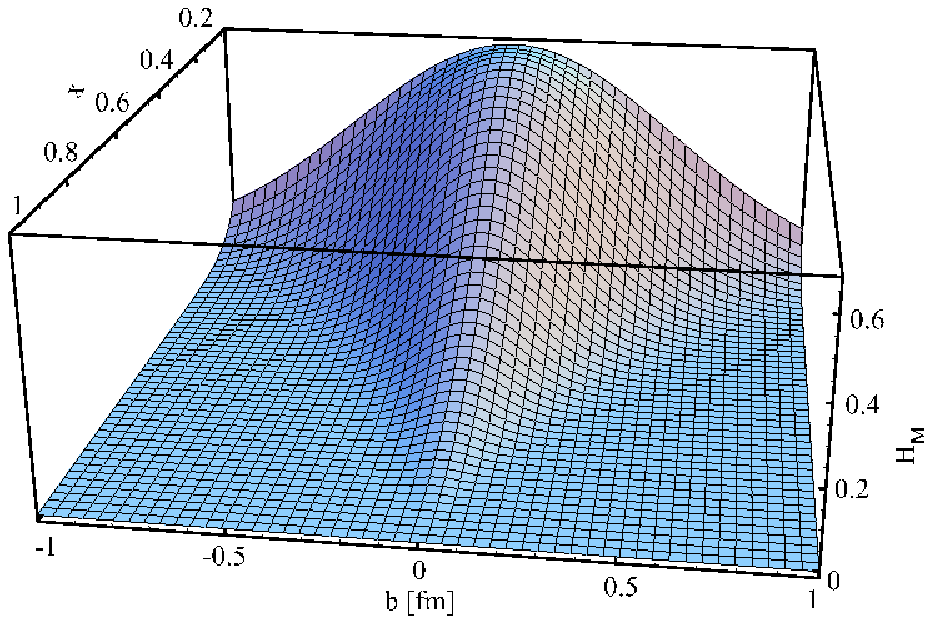}
\caption{Top: The red curve represents the form factor $G^*_M$ 
obtained by~\cite{pascalutsa} using a Regge like parameterization of 
the elastic isovector form factor applied to the $N\to \Delta$ transition. 
Bottom: The distribution of the transverse impact parameter 
$\vec b_\perp$ and longitudinal imomentum. \label{Pascalutsa_GPD}}
\end{figure}

The first practical application of GPDs to resonances were reported in
\cite{stoler} for the $N\to\Delta$ transition. It was shown that the
anomalously rapid falloff of the  $G^*_M$ can be directly related to
the  unexpectedly rapid falloff of the elastic helicity flip
$F_2$, which had been  recently discovered, by constraining the $N\to\Delta$ GPD by the isovector
part of the elastic scattering form factors. Figure~\ref{Pascalutsa_GPD} 
shows a more recent~\cite{pascalutsa} fit to $G^*_M$, which was
obtained from GPDs constrained from elastic scattering using a Regge
like parameterization. 
The Fourier transform of the GPD gives the  distribution of the impact parameter ~\cite{burkardt} in the transverse plane vs. the longitudinal momentum fraction, i.e. $ H_M(\vec  b_\perp, x)$, also shown in Fig. ~\ref{Pascalutsa_GPD} .

$$ H_M(x,\vec b_\perp)= \int { {d^2(\vec q _\perp)}\over{2\pi^2} } e^{i(\vec b_\perp \cdot \vec q_\perp)} H_M (-\vec q_\perp^2,x,0).$$

\vspace{0.1in}
The application of the GPD formalism to nucleon excitation is most readily seen in the $J=1/2 \to 1/2$ transitions such as the 
$N\to P_{11}(1440)$ or the $S_{11}(1535)$.  

\vspace{0.1in}

 \noindent{\bf \large The $\bf N \to  P_{11}$(1440):}
 
Since the $N\to P_{11}(1440)$ is a $1/2^+ \to1/2^+$ transition, its current structure is similar to elastic scattering, i.e.

$$\Gamma ^\mu _{P11}  =  {{F_1^{P11}(q^2 )}\over {M^2_N} }\left( {q^2\gamma ^\mu   - \not q q^\mu  } \right)  + \frac{{F_2^{P11} (q^2 )}}{{2M_N}}i\sigma ^{\mu \nu } q_\nu $$

\noindent  which immediately leads to  a GPD structure and related form factors  exactly as in elastic scattering:

\begin{eqnarray}
\nonumber
\frac{P^+}{{2\pi }}\left. {\int {dy^ -  } e^{ix\bar P^ +  y^ -  } \left\langle {P_{11}(p')} \right|\bar \psi _{P11}\left( { - y/2} \right)\gamma^\nu n_\nu\psi \left( {y/2} \right)\left| {N(p)} \right\rangle } \right|_{y^ +   = \vec y_ \bot   = 0}  
&=&\nonumber\\
H_{P11} \bar u(p'){{\left( {q^2\gamma ^\mu   - \not q q^\mu  } \right)n_\mu}\over{M^2_N}}u(p) + E_{P11} \bar u(p')i\sigma ^{\mu \nu } \frac{{n_\mu q_\nu  }}{{2M_N }}u(p) 
\end{eqnarray}

$$F_{1P_{11} }^q (t) = \int { H_{P_{11} }^q (x,\xi ,t)} dx\,\,\,\,\,F_{2P_{11} }^q (t) = \int { E_{P_{11} }^q (x,\xi ,t)} dx$$
 
\vspace{0.1in}

 \noindent{\bf The N $\to$ S$_{11}$(1535):}
The  $S_{11}(1535)$ has $J^\pi = 1/2^-$ and is the chiral negative parity partner of the nucleon.  The current structure has an extra $\gamma_5$ and can be written

$$\Gamma ^\mu _{S11}  =  {{F_1^{S11}(q^2 )}\over {M^2_N} }\left( {q^2\gamma ^\mu   - \not q q^\mu  } \right)\gamma_5  + \frac{{F_2^{S11} (q^2 )}}{{2M_N}}i\sigma ^{\mu \nu } q_\nu \gamma_5$$

\noindent which leads to 

\begin{eqnarray}
\label{GPD-S11}
\nonumber
\frac{P^+}{{2\pi }}\left. {\int {dy^ -  } e^{ix\bar P^ +  y^ -  } \left\langle {S_{11}(p')} \right|\bar \psi  _{S11}\left( { - y/2} \right)\gamma^\nu n_\nu\psi \left( {y/2} \right)\left| {N(p)} \right\rangle } \right|_{y^ +   = \vec y_ \bot   = 0} 
&=&\nonumber\\
H_{S11} \bar u(p'){{\left( {q^2\gamma ^\mu   - \not q q^\mu  } 
\right)n_\mu}\over{M^2_N}}\gamma _5 u(p) + E_{S11} \bar u(p')i
\sigma ^{\mu \nu } \gamma _5 \frac{{ q^\nu n_\mu }}{{2M_N }}u(p) 
\end{eqnarray}

\noindent with

$$F_{1S_{11} }^q (t) = \int { H_{S_{11} }^q (x,\xi ,t)} dx\,\,\,\,\,F_{2S_{11} }^q (t) = \int { E_{S_{11} }^q (x,\xi ,t)} dx$$

\vspace{0.1in}
\noindent{\bf The N $\to$ $ \bf \Lambda$, $ \bf \Sigma$:}
Hard exclusive processes with strangeness production was treated in Refs. ~\cite{frankfurt}
and  ~\cite{goeke},  in which effects related to  $SU(3)$ flavor 
symmetry-breaking are discussed. The GPDs correspond to the process where a 
non-strange quark  is taken out of the initial nucleon at the space-time 
point $y/2$, and then a strange quark is put back exciting a hyperon at the 
space-time point  $y/2$. Following Ref.~\cite{frankfurt} the strangeness changing  distributions for $N \to  \Sigma, \Lambda$ transitions:

\begin{eqnarray}
\label{GPD-Y}
\nonumber
\frac{P^+}{{2\pi }}\left. {\int {dy^ -  } e^{ix\bar P^ +  y^ -  } \left\langle {Y(p')} \right|\bar\psi\left( { - y/2} \right)\bar a_s \left( { - y/2} \right) \gamma^\nu n_\nu a_q \left( {y/2} \right)\psi\left( {y/2} \right) \left| {N(p)} \right\rangle } \right|_{y^ +   = \vec y_ \bot   = 0} 
&=&\nonumber\\
H_{N\to Y} \bar u(p'){{\left( {q^2\gamma ^\mu   - \not q q^\mu  }
\right)n_\mu}\over{M^2_N}}\gamma _5 u(p) + E_{N\to Y} \bar u(p')i\sigma ^{\mu
\nu } \gamma _5 \frac{{ q^\nu n_\mu }}{{2M_N }}u(p) 
\end{eqnarray}

\noindent where $\bar a_s$   is the creation operator of a strange quark and 
$\bar a_q$ the annihilation operator of a non-strange quark, $u$ or $d$, at $-y/2$ and\  $y/2$,  respectively. 

\vspace{0.3in}
\noindent{\bf Light Cone Sum Rules (LCSR).}
\vspace{0.1in}

The Light-Cone Sum Rule (LCSR) approach allows one to calculate form factors 
(and, potentially, also GPDs) using much more limited information compared  to
the full nonperturbative wave functions, albeit with some assumptions. The groundwork of the method of LCSRs was laid  in ~\cite{LCSR}. 
Following the work \cite{Braun:2001mu,Braun:2006hz} devoted to electromagnetic nucleon form factors 
the $N \to \Delta$ transitions were considered in \cite{Braun:2005be1} 
and the $N\to S_{11}$  transitions in \cite{braun-S11}. 
LCSRs have also been applied to threshold pion electroproduction \cite{Braun:2006td1,Braun:2007pz1}.

For example, in order to calculate the transition amplitude 
\begin{equation}
\label{LCSR-S11}
\left\langle {N_{S11} \left( {P'} \right)\left| {j_\mu ^{em} } 
\right|N\left( P \right)} \right\rangle  = 
N_{S11} \left( {P'} \right)\left( {F_1^{S11} (q^2){{\left( {q^2\gamma _\mu   - \not q q_\mu  } \right)}
\over{M^2_N}} F_2^{S11} (q^2)i\sigma _{\mu \nu } \frac{{\,\,q^\nu  }}{{2M_N }}} \right)
\gamma _5 N\left( P \right)
\end{equation}
one considers  the correlation function 
$$\int\! dx\, e^{-iqx}\langle S11(P')| T \{ \eta (0) j_\mu^{\rm em} (x) \} | 0 \rangle $$ in which
$\eta$ is a suitable operator with nucleon quantum numbers.
A popular choice is, for example
\begin{equation}
  \eta(0) = \epsilon^{ijk}\! \left(u_i C\gamma_{\mu} u_j\right)\!(0) \!\gamma_5\gamma^{\mu} d_k(0)
\end{equation}   
where $u_{i,j}(0)$ and $d_k(0)$ are u-quark and d-quark field operators, $i,j,k = 1,\ldots,3$ is 
the color index; $C$ is the charge conjugation matrix. The electromagnetic current is
\begin{equation}
  j_\mu^{\rm em} (x) = e_u \bar u(x)\gamma_\mu u(x) + e_d \bar d(x)\gamma_\mu d(x) 
+ e_s \bar s(x)\gamma_\mu s(x) 
\end{equation}

Making  use of the duality of QCD quark-gluon and hadronic degrees of freedom  
through dispersion relations one can write a representation for the 
form factors appearing in (\ref{LCSR-S11}) in terms of the distribution amplitudes 
(DAs) of the $S_{11}$ resonance.  These DAs  correspond to the momentum fraction distributions 
of the three quarks in the $S_{11}$ at small transverse separations. Unlike the $S_{11}$ wave functions
themselves, the DAs can be accessed through lattice calculations.

The leading twist-3 nucleon (proton) DA can be defined as a matrix element of the 
nonlocal light-ray operator that involves quark fields of given helicity    
$q^{\up(\down)} = (1/2) (1 \pm \gamma_5) q$

\begin{eqnarray}\label{eq_nonlocal}
\label{vector-twist-3}
\lefteqn{
\langle 0 | 
\epsilon^{ijk}\! \left(u^{\up}_i(a_1 n) C \!\!\not\!{n} u^{\down}_j(a_2 n)\right)  
\!\not\!{n} d^{\up}_k(a_3 n) 
|N(P)\rangle =}
\nonumber\\
&=& - \frac12 f_N\,pn\! \not\!{n}\, u_N^\up(P)\! \!\int\! [dx] 
\,e^{-i pn \sum x_i a_i}\, 
\varphi_N(x_i)\,.
\end{eqnarray}
Here $P_\mu$, $P^2=m_N^2$, is the proton momentum,  $u_N(P)$ is the usual Dirac spinor in 
relativistic normalization, $n_\mu$ an arbitrary light-like vector  with $n^2=0$, as defined above, and $C$
the charge-conjugation matrix. The variables $x_1,x_2,x_3$ have 
the meaning of the momentum fractions carried by the three valence quarks and the 
integration measure is defined as 
$\int [dx] = \int_0^1  dx_1 dx_2 dx_3 \delta(\sum x_i-1)$.
The Wilson lines that ensure gauge invariance are inserted between the quarks;
they are not shown for brevity.

The definition in (\ref{eq_nonlocal}) is equivalent to the following
form of the valence proton state
\begin{equation}
| p,\uparrow \rangle = f_N \int \frac{[dx]\, \varphi_N(x_i)}
                                    {2\sqrt{24x_1x_2x_3}}
\left \{
|u^\uparrow(x_1)u^\downarrow(x_2) d^\uparrow(x_3)\rangle
-
|u^\uparrow(x_1)d^\downarrow(x_2) u^\uparrow(x_3)\rangle \right \},
\end{equation}
where the arrows indicate the helicities and
the standard relativistic normalization for the states and
Dirac spinors is implied.

The nonlocal operator appearing on the l.h.s. of (\ref{vector-twist-3})
does not have a definite parity. Thus the same operator couples also 
to $N^*(1535)$ and one can define the corresponding leading-twist DA as

\begin{eqnarray}
\label{vector-twist-3-star}
\lefteqn{
\langle 0 | \epsilon^{ijk}\! 
\left(u^{\up}_i(a_1 n) C \!\!\not\!{n} u^{\down}_j(a_2 n)\right)  
\!\not\!{n} d^{\up}_k(a_3 n) |N_{S11}(P)\rangle =}
\nonumber\\
&=&  \frac12 f_{N^*}\,pn\! \not\!{n}\, u_{S11}^{\up}(P)\! \!\int\! [dx] 
\,e^{-i pn \sum x_i a_i}\, 
\varphi_{S11}(x_i)\,
\end{eqnarray}
where, of course, $P^2=m_{N^*}^2$.
The normalization constants $f_N$, $f_{N^*}$ are defined such that 
the DAs are normalized to unit integral:
\begin{equation}
  \int [dx]\, \,\varphi(x_i) = 1\,.
\end{equation}

As an example of the results of the LCSR calculation of the helicity amplitides
$A_{11}$ and $S_{11}$ obtained  by \cite{braun-S11,nstar} using the lattice QCD estimates of the
relevant distribution amplitudes \cite{Warkentin:2008iu}, is shown in Figs 1 and 2 of the contribution of V. Braun 
~\cite{Braun-WP}  to these proceedings. 

To close the circle, one can relate the LCSR matrix element to the GPDs: 

\begin{equation}\label{LCSR-GPD}
\left\langle {N_{S11} \left( {P'} \right)\left| {j_\mu ^{em} } \right|N\left( P \right)} \right\rangle \sim
\int dx H_{S11} \bar u(p'){{\left( {q^2\gamma ^\mu   - \not q q^\mu  } \right)}\over{M^2_N}}\gamma _5 u(p) + \int dx E_{S11} \bar u(p')i\sigma ^{\mu } \gamma _5 \frac{{ q^\nu }}{{2M_N }}u(p) 
\end{equation}

%\end{document}

%\newpage
\section{ Constituent Quark Models }
%\begin{center}
%%{\bf Relativistic Constituent Quark Models}
%%\\
%{\large S. Capstick, M. Giannini, E. Santopinto, Q. Zhao, B. Zou}
%\end{center}

%\section{Relativistic Constituent Quark Models}
The study of hadron properties can be performed within a microscopic
approach based on quark degrees of freedom and their interactions. The
widely accepted framework is provided by Quantum ChromoDynamics (QCD),
which is of course fully relativistic, however it is usable only in
particular conditions, mainly at high momentum transfer.  There are
now many important results in the Lattice QCD (LQCD) but the present
computer capabilities do not yet allow to extract all the hadron
properties in a systematic way. In the meanwhile, one can rely on
models, eventually based on QCD or LQCD.  An important class of such
models is provided by Constituent Quark Models (CQM), in which quarks
are considered as effective internal degrees of freedom and can
acquire a mass and even, in certain approaches, a finite size.  There
are many versions of CQM, which differ according to the chosen quark
dynamics: one-gluon exchange and a three-body force \cite{ho,ho1,rm_ho},
algebraic \cite{bil}, hypercentral (hCQM) \cite{pl}, Goldstone Boson
Exchange (GBE) \cite{olof},instanton \cite{bn}.  In most cases they
have been applied to the description of many hadron properties
(spectrum, elastic form factors, transition form factors,...) and have
also been relativized.

The construction of a Relativistic Constituent Quark Model (RCQM)
means a) the use of a relativistic kinetic energy for the quarks; b)
the application of Lorentz boosts in order to describe baryons in
motion; c) the formulation of quark dynamics within a relativistic
hamiltonian using one of the forms introduced by Dirac: front, instant
or point form which provide different realizations of the Poincar\'{e}
group. Of course c) implies also both a) and b). An alternative way of
building a relativistic baryon description is given by a
Bethe-Salpeter approach (BS) \cite{bn,dsm-1}. As far as the spectrum is
concerned, a) is often the only relativistic aspect which is
considered, however, in electron scattering the recoil of the struck
nucleon becomes relativistic as the momentum transfer
increases. However, when baryon resonances are excited, such effects may 
be softened because of the higher mass of the recoiling resonance
\cite{mds2}.

\begin{figure}[htb]
\includegraphics[width=0.49\textwidth]{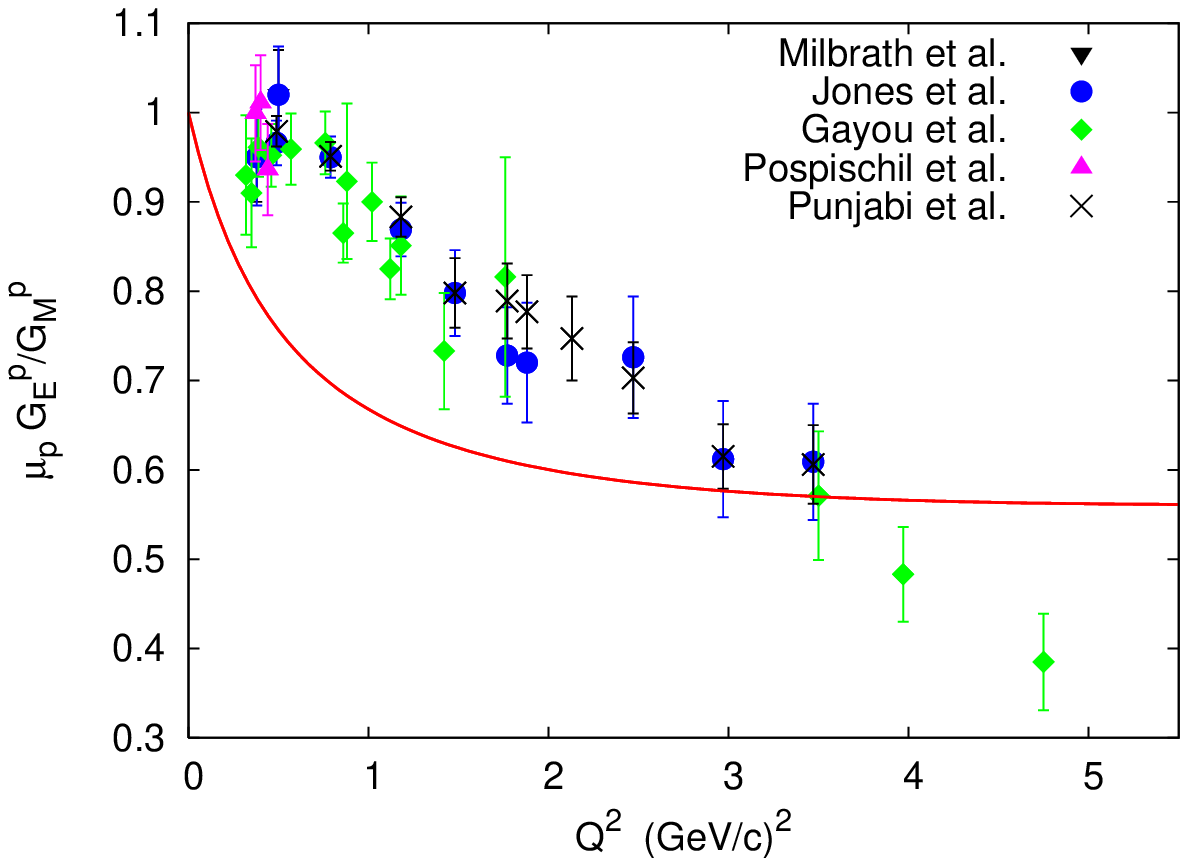}
\includegraphics[width=0.49\textwidth]{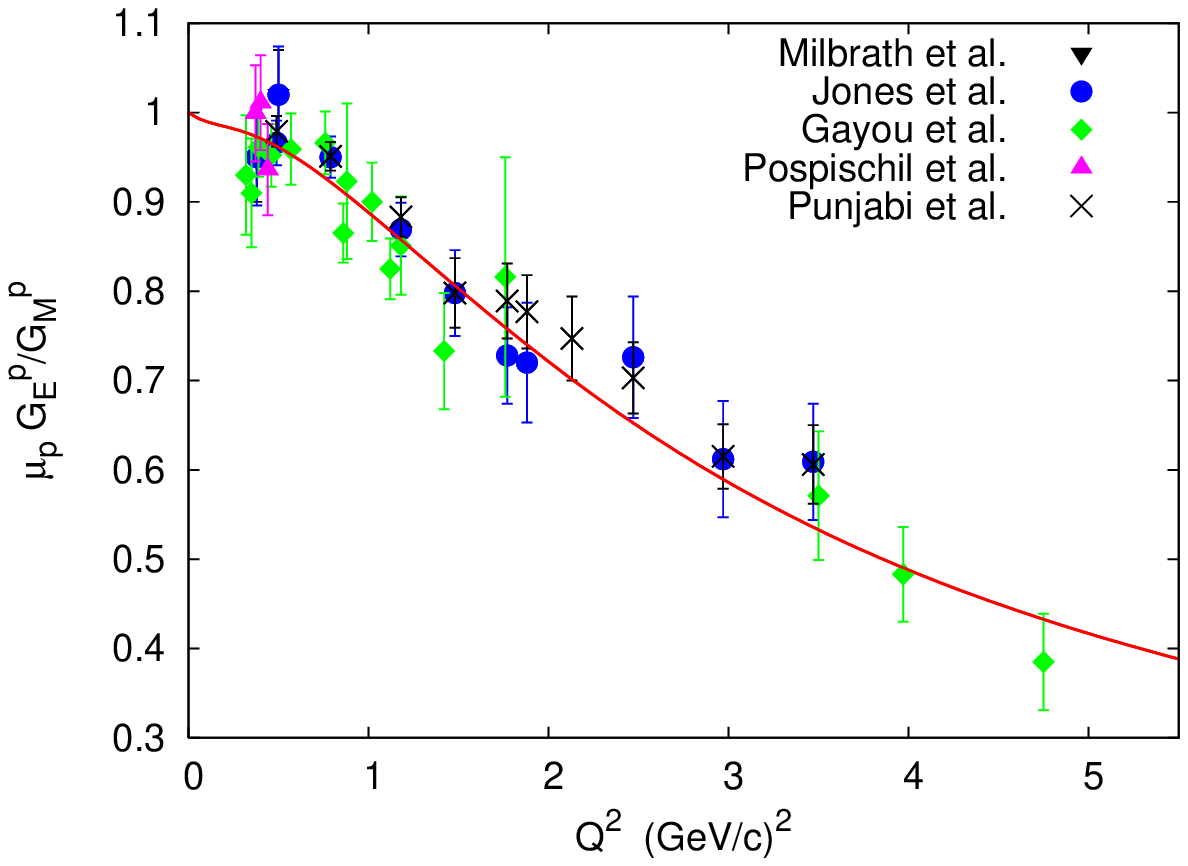}
\caption{The ratio $G_E^p/\mu_p G_M^p$ calculated with the hCQM 
\cite{ff_rel}, without (left) and
with (right) constituent quark form factors. Data are in ref. \cite{ff_rel}.}
\label{fig:cqm_ff}
\end{figure}

There are now many results obtained with relativistic Constituent
Quark Models (RCQM). The relativized h.o. with light front has been
applied to the calculation of the elastic nucleon form factors and of $\gamma_{v}NN^*$ helicity amplitudes 
\cite{Az93,cqm-rel-1,cqm-rel-2,cqm-rel-4,rm_ho,rm_ha}. A good
description of the nucleon elastic form factors is obtained both in
the GBE model in the point and front forms \cite{boffi,melde} and in the instanton BS
approach \cite{bn}. The hCQM has been used for a systematic prediction
of the helicity amplitudes \cite{cqm-rel-3} (although in its non
relativistic version) and of the elastic form factors in a fully
relativistic formulation using the point form \cite{ff_rel}. Comparing
these predictions with the helicity amplitudes data \cite{cqm-rel-3}, one
observes a systematic lack of strength, which, according to a wide
consensus, is ascribed to the missing $q\bar{q}$ pairs in the
outer region \cite{brag07}. For medium $Q^2$ the behaviour is fairly
well reproduced, although some discrepancies arise, probably because
relativity is not taken into account completely \cite{mds2}. On the
other hand, the prediction for the elastic form factors is close to
the data, but a very good fit is obtained by introducing quark form
factors (Fig. ~\ref{fig:cqm_ff}).

The two aspects just mentioned, the $q\bar{q}$ pair effects and
the quark form factors presumably will play a key role in the
description of the baryon excitation in the $Q^2$ range accessible
with 12 GeV electrons. This is certainly a transition region between
the phase where the CQ with mass and size are the dominant degrees of
freedom and the range where the asymptotic behaviour dominated by
current quarks starts to be effective. The presence of $q\bar{q}$
(or meson cloud) effects points towards the unquenching of the quark
models \cite{brag07}. This problem has been addressed for the meson
sector within the flux tube model \cite{gi} but recently also the
baryon sector has been studied \cite{bs}. With the availability of
unquenched CQM, it will be possible to describe the microscopic
mechanisms leading to the excitation of resonances and the production
of mesons. Therefore both the electromagnetic and strong decay of the 
resonances
will be described in a consistent way. The description of the spectrum 
will be
also more realistic. In fact, in all the presently used CQM, the energies 
of
the baryon resonanances are sharply defined, while in an unquenched 
approach
the excited states acquire a non zero width thanks to the coupling with 
the
continuum.

The meson cloud effects are certainly relevant in the low
$Q^2$ region \cite{lt}. A calculation with a dynamical model \cite{dmt} 
shows
that actually the contribution of the pion cloud to the helicity 
amplitudes
decreases with increasing $Q^2$ and seems to partially compensate the lack 
of
strength of the CQM calculations. However, with increasing momentum 
transfer
the excitation of resonances will also allow testing of the short distance
behaviour of the $q\bar{q}$ production mechanism and, in particular, of 
the meson
production.

The availability of high-intensity electron beams at 12 GeV will also
allow the probing at high $Q^2$ of baryons that have a more
complicated structure than a simple three quark configuration. If
multiquark configurations, such as $qqqq\bar{q}$, can mix with the
conventional three-quark components of a baryon, they may have a
different $Q^2$-dependence compared with the $qqq$ component \cite{cqm-seaq,5q_comp2}.
The phenomenological quark form factors which have been introduced up
to now contain and mix contributions from both the structure of the 
effective
(consituent) quarks and from the dynamics not explicitly  included in CQM, 
such
as the
$q\bar{q}$ pair creation or meson production effects. By unquenching the 
CQM, it will
be possible to disentangle the quark form factors and test the onset of 
the transition
to the asymptotic QCD current quarks.

In absence of a consistently unquenched approach, one can use the CQM in order 
to
provide constraints on the parameters describing the leading order
baryon-photon and baryon-meson vertices by considering explicit 
quark-photon and
quark meson couplings. In exclusive meson production channels, an economic 
way to include a complete set of intermediate baryon resonances is the
introduction of effective Lagrangians for the constituent-quark-meson
couplings. One can then explicitly construct  transition operators and by 
studying their
$Q^2$ evolution one can establish relations
between the internal quark motions and EM and strong form factors in
exclusive meson electroproduction reactions. Various quark model
approaches for the reaction process can be tested and compared with
systematic experimental measurements.

Quark-hadron duality has been one of the most striking phenomena
observed in electron-proton inclusive scattering, where the
low-energy exclusive resonance excitations are related to the
high-energy inclusive scaling behaviour through a local average over
the resonance structure functions. Recent experimental data from
JLab have tested this empirical phenomenon to high precision and
initiated renewed interest in this field. In particular, the idea of 
quark-hadron
duality has been used in a recent analysis \cite{PSR} which allowed to 
identify objects
inside the proton having a finite constituent size and non-zero form 
factors.
The role of quark-hadron duality has been investigated also in exclusive 
meson
photoproduction, where a restricted locality of quark-hadron duality was
shown to be important \cite{zc} and related to deviations from the pQCD 
counting rules
above the resonance region. In the quark model framework, the resonance 
phenomena are
dual to the quark motion correlations and the study of vector meson photo- 
and
electroproduction from low to high $Q^2$ is expected to allow 
an interesting test of this phenomenon, and shed light
on the transition between the perturbative and strong interaction
regimes of QCD.

Therefore, the excitation at high $Q^2$ of resonances may provide new
information concerning the fundamental underlying QCD mechanisms
responsible for the baryon structure and quark confinement.

To conclude, the NRCQM has provided a consistent framework for the
description of a large number of hadron properties. However, in order
to be applcable to the high $Q^2$ regime, the CQM not only has to be formulated in a consistent relativized
framework, according to the methods
mentioned above, but it should also include another fundamental relativistic
requirement,
that of the possibility of the creation of quark-antiquark pairs.

%\end{document}

%\newpage
\section{ Status of JLab Data Analysis }
%\documentclass[prc,superscriptaddress,amssymb,amsmath,amsfonts,aps]{revtex4}
%\setlength{\topmargin}{-1.0cm}
%\usepackage{graphicx}
%\usepackage{dcolumn}
%\usepackage{epsfig}
%\begin{document}

%\begin{center}
%{\it Jefferson Lab, 12000 Jefferson Ave, Newport News, VA 23606, USA}
%\end{center}
%\affiliation{Jefferson Lab, 12000 Jefferson Ave, Newport News, VA 23606, USA} 

%\noaffiliation

%\thispagestyle{empty}

%\maketitle

%\begin{center}
%%\section{Phenomenological Analysis of Single and Charged Double Pion Electroproduction at JLAB}

%\large{I.~G. Aznauryan and V.~I.~Mokeev}

%%{\it Jefferson Lab, 12000 Jefferson Ave, Newport News, VA 23606, USA}
%\end{center}

\vspace{0.1in}
\begin{center}
{\bf $N^{*}$ studies in meson electroproduction with CLAS}
\end{center}
\vspace{0.1in}

The comprehensive experimental data set obtained with the CLAS detector on
single pseudoscalar meson electroproduction, e.g. $p\pi^0$,
$n\pi^{+}$, $p\eta$, and $K\Lambda$
\cite{Joo02,Ungaro:2006df,Joo03,Joo04,Joo05,egiyan06,park08,thomp01,den07,carman03,ambroz07,devita02,biselli03,biselli06}
and double charged pion electroproduction \cite{Ri03,Fe07,Fe007,Fe07a} opens
up new opportunities for studies of the $\gamma_{v}NN^*$ transition helicity
amplitudes (i.e. the $N^{*}$ electrocoupling parameters)
\cite{Burk07,Burk05a,Lee04a}. The CLAS data for the first time provided
information on many observables in these exclusive
channels, including fully integrated cross sections and a variety of
1-fold differential cross sections complemented by single and double
polarization asymmetries in a range of $Q^2$ from
0.2 to 4.5 GeV$^2$. This comprehensive information makes it
possible to utilize well established constraints from dispersion
relations and to develop phenomenological approaches in order to determine the
$Q^2$-evolution of the $N^{*}$ electrocoupling parameters by fitting them
to all available observables combined.  Several
phenomenological analyses of the experimental data on single (1$\pi$) and charged double
pion (2$\pi$) electroproduction have already
been carried out within the CLAS Collaboration
\cite{Az03b,Az05d,Az06,Az08g,Mo06-1,Mo06,Mo08m,Mo07}. They allowed us to determine
transition helicity amplitudes and the corresponding transition form
factors for a variety of low lying states: $P_{33}(1232)$,
$P_{11}(1440)$, $D_{13}(1520)$, $S_{11}(1535)$ at photon virtualities
from 0.2 to 4.5 GeV$^2$. Typical examples for resonance
electrocoupling parameters are shown in Fig.~\ref{p11d13}.
The 2$\pi$ data enhance substantially our capabilities  for the studies 
of $N^{*}$ with masses above 1.6 GeV. Many of these resonances decay
predominantly to N$\pi\pi$ final states.
The analysis of 2$\pi$ data at W $>$ 1.6 GeV allowed us for the
first time to map out the $Q^{2}$ evolution of electrocoupling
parameters for resonances with masses above 1.6 GeV that
preferably decay by 2$\pi$ emission: $S_{31}(1620)$, $D_{33}(1700)$
and $P_{13}(1720)$ \cite{Mo06,Mo09ch}. In analysis of the 2$\pi$ electro 
production data \cite{Ri03} we observed a signal
from a $3/2^{+}(1720)$ candidate state whose quantum numbers and
hadronic decays parameters are determined from the fit to the measured
data.

 There are
up to three transition helicity amplitudes $A_{1/2}(Q^{2})$,
$A_{3/2}(Q^{2})$, and $S_{1/2}(Q^{2})$, that fully describe the excitation
of a resonance by virtual photons. Resonance excitations may
also be described in terms of $F_{1}^{*}(Q^2)$, $F_{2}^{*}(Q^2)$ or 
$G_{E}^{*}(Q^2)$, $G_{M}^{*}(Q^2)$ transition form factors (for states
with spin~$>$ 1/2 we also have a third form factor in both
representations), that are  used in the electromagnetic $N
\rightarrow N^{*}$ transition current. They play a similar role as the
elastic form factors. The descriptions of resonance excitations by
transition form factors or transition helicity amplitudes are
equivalent and can be uniquely expressed in terms of each other
\cite{Az08f}. They can be determined either by fitting resonance parts of
production amplitudes within
the framework of a Breit-Wigner ansatz \cite{So71} or by applying
various multi-channel resonance parameterizations \cite{Vrana}.
 
Full production amplitudes in all meson electroproduction channels represent a 
superposition of resonant contributions and complicated non-resonant processes. 
In order to determine the $N^{*}$ electrocoupling parameters a reliable
separation of resonant and non-resonant parts contributing to the
meson electroproduction amplitudes is needed. This is one of the most
challenging problems for the extraction of $N^*$ electrocoupling
parameters. The amplitudes of effective meson-baryon interactions in
exclusive electroproduction reactions cannot be expanded in a small
parameter over the entire resonance region. It is impossible to select
contributing diagrams through a perturbative expansion. So far, no
approach has been developed that is based on a fundamental theory and
that would allow either a description of an effective meson-baryon
Lagrangian or a selection of the contributing meson-baryon mechanisms 
from first principles. We therefore have to rely on fits to the
comprehensive experimental data of various meson electroproduction
channels from CLAS to develop reaction models that contain the
relevant mechanisms. This approach allows us to determine all the essential 
contributing
mechanisms based on their manifestations in the kinematic dependencies of
measured observables. 

Nucleon resonances have
various decay modes and hence  manifest themselves in different meson
electroproduction channels. Contributions of non-resonant amplitudes
are substantially different in the different meson electroproduction
channels \cite{Lee04a,Pe02}. On the other hand, the $N^*$
electrocoupling parameters remain the same in all meson
electroproduction channels. They are fully determined by the
$\gamma_vNN^*$ vertices and independent from the hadronic decay of
the resonance.  The successful description of a large body of 
observables in various exclusive channels with a common set of $N^{*}$
electrocoupling parameters gives evidence that the $\gamma_vNN^*$ helicity 
amplitudes can be reliably determined from different hadronic final states. 
In the future, this
analysis will be carried out in a complete coupled channel approach
which is currently being developed at EBAC \cite{Lee06,Lee07a,Lee07}.  

1$\pi$ and 2$\pi$ electroproduction are the two dominating exclusive channels 
in the resonance region. The 1$\pi$ exclusive channel is mostly sensitive 
to $N^{*}$'s with masses lower than 1.65~GeV. Many resonances of
heavier masses decay predominantly by two pion emission. Thus the 2$\pi$
exclusive channel offers better opportunities to study the electrocoupling
parameters of these high-lying states. The final states in 1$\pi$ and 2$\pi$
channels have considerable hadronic interactions. The cross section for 
the $\pi N
\rightarrow \pi \pi N$ reaction is the second largest of all of the
exclusive channels for $\pi N$ interactions. Therefore, for $N^*$
studies both in single and double pion electroproduction, information 
on the mechanisms contributing to each of these channels is needed in
order to take properly into account the impact from coupled-channel
effects on the exclusive channel cross sections. The knowledge of
single and double pion electroproduction mechanisms becomes even more
important for $N^*$  studies in channels with smaller cross sections
such as $p\eta$ or $K\Lambda$ and $K\Sigma$ production, as they can
be significantly affected in leading order by coupled-channel effects
produced by  their hadronic interactions with the dominant single and
double pion electroproduction channels. Comprehensive studies of single and 
double pion electroproduction are of key
importance for the entire baryon resonance research program.

\begin{figure}[ht]
\includegraphics[width=14.cm]{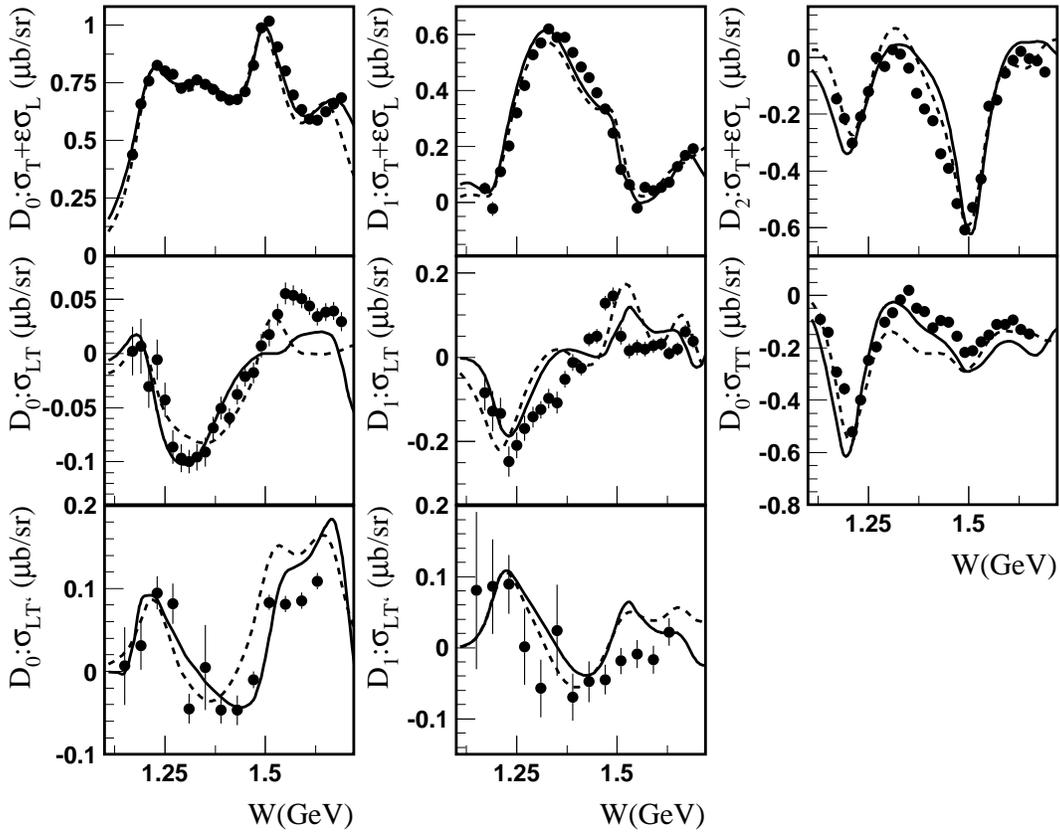}
\caption{The results for the Legendre moments of the
$\vec{e}p\rightarrow en\pi^+$ structure functions
in comparison with experimental data \cite{park08} for 
$Q^2=2.44~$GeV$^2$.
The solid (dashed) curves
correspond to the analyses made using DR (UIM) approach.
}
\label{im1m}
\end{figure}

\vspace{0.1in}
\begin{center}
{\bf Analysis approaches for the single meson electroproduction data}
\end{center}
\vspace{0.1in}

Over the past 40 years, our knowledge of electromagnetic excitations
of nucleon resonances was mainly based on single pion photo- and
electroproduction. These reactions have been the subject of extensive
theoretical studies based on dispersion relations and isobar models.
The dispersion relation (DR) approach has been developed on the basis
of the classical works \cite {Chew57,Fubini58} and played an extremely
important role in the extraction of the resonance contributions from
experimental data. Dispersion relations provide stringent constraints on 
the real part of the reaction
amplitudes that contain the most significant part of the non-resonant 
contributions. Starting in the late 1990's the Unitary
Isobar Model \cite{Dr99} (also known as MAID), became widely used for
the description of single-pion photo- and electroproduction
data. Later this approach has been modified \cite{Az03b} by
incorporating Regge poles to describe the high energy regime. 
This extension of the isobar model
enables a good description of all photo-production multipole
amplitudes with angular momenta $l\,\leq\,3$ up to an invariant mass
$W=2$~GeV using a unified Breit-Wigner parametrization of the
resonance contributions in the form as proposed by Walker
\cite{Wa69}. Dispersion relations and the Unitary Isobar Model (UIM)
\cite{Az03b} have been successfully used for the analysis
\cite{Az05d,Az06,Az08g} of the CLAS
\cite{Joo02,Joo03,Joo04,egiyan06,Ungaro:2006df,park08} and the world data 
to extract resonance electrocouplings from the data on
cross sections and longitudinally polarized electron beam asymmetries for
the reactions $p(\vec{e},e'p)\pi^0$ and $p(\vec{e},e'n)\pi^+$ in the
first and second resonance region. The quality of these results is
best characterized by the following $\chi^2$ values: $\chi^{2}<1.6$ at 
$Q^{2}=0.4$ and $0.65$~GeV$^2$ and $\chi^{2} < 2.1$ at 
$1.7 < Q^2 < 4.5$~GeV$^2$. In the analyses
\cite{Az03b,Az06,Az05d,Az08g}, the $Q^2$ evolution of the electrocoupling 
amplitudes for the lower-lying resonances with $W < 1.6$~GeV 
have been established for $Q^2$s up to $4.5$~GeV$^2$. The
comparison of two conceptually different approaches, DR and UIM,
allows us to conclude that the model-dependence of the
obtained results is relatively small.

The background in both approaches, DR and UIM, contains Born terms
corresponding to $s$- and $u$-channel nucleon exchanges and the
$t$-channel pion contribution, and thus depends on the proton,
neutron, and pion form factors. The background of the UIM contains
also the $\rho$ and $\omega$ $t$-channel exchanges, and thus
contributions of the form factors
$G_{\rho(\omega)\rightarrow\pi\gamma}(Q^2)$. The proton magnetic and
electric form factors as well as the neutron magnetic form factor are 
known from the existing experimental data, for $Q^2$ up to 32, 6, and
10~GeV$^2$, respectively
\cite{Be71,Pr71,Ba73,Sti93,Wa94,An94,Jo00,Ga01,Lu93,Br05}. This
information on the proton and neutron elastic form factors combined with
the parametrization of the proton electric form factor from
polarization experiments \cite{Ar07} can be readily used for the
analysis of the pion electroproduction data up to quite large values
of $Q^2$. The neutron electric form factor, $G_{E_n}(Q^2)$, is measured
up to $Q^2=1.45$~GeV$^2$ \cite{Madey:2003av}. A parametrization of
all existing data on  $G_{E_n}(Q^2)$ \cite{Madey:2003av} can be used to
extrapolate $G_{E_n}(Q^2)$  to higher four momentum transfers. The
pion form factor $G_{\pi}(Q^2)$ has been studied for $Q^2$ values from
$0.4$ to $9.8$~GeV$^2$ at CEA/Cornell \cite{Be76,Be78} and more
recently at JLab \cite{Ho06,Ta07}. All these measurements show that
the $Q^2$ dependence of $G_{\pi}(Q^2)$ can be described by a simple
monopole form $1/(1+\frac{Q^2}{0.46~{\rm GeV}^2})$ \cite{Be76,Be78} or 
$1/(1+\frac{Q^2}{0.54~{\rm GeV}^2})$ \cite{Ho06,Ta07}, respectively. 
There are no measurements on the 
$G_{\rho(\omega)\rightarrow\pi\gamma}(Q^2)$ form factors. However,
investigations, one based on QCD sum rules \cite{El84} and another one
on a quark model \cite{Az90}, predict that the $Q^2$ dependence of
these form factors follows closely the dipole form. Therefore our
corresponding background estimations proceed from the assumption that
$G_{\rho(\omega)\rightarrow\pi\gamma}(Q^2)\sim
1/(1+\frac{Q^2}{0.71~{\rm GeV}^2})^2$.

In figure~\ref{im1m} we present as an example, 
the description  
of $\vec{e}p\rightarrow en\pi^+$  data \cite{park08}
for one specific $Q^2$ value.
The results are shown in terms of the
Legendre moments of structure functions. This allows
us to compare the analysis results with  
experimental data for all energies and angles.

\vspace{0.1in}
\begin{center}
{\bf Meson-baryon model approach JM for the 2$\pi$ electroproduction analysis}
\end{center}
\vspace{0.1in}
  
\begin{figure*}[ht]
\begin{center}
%\hspace{-4cm}
\epsfig{file=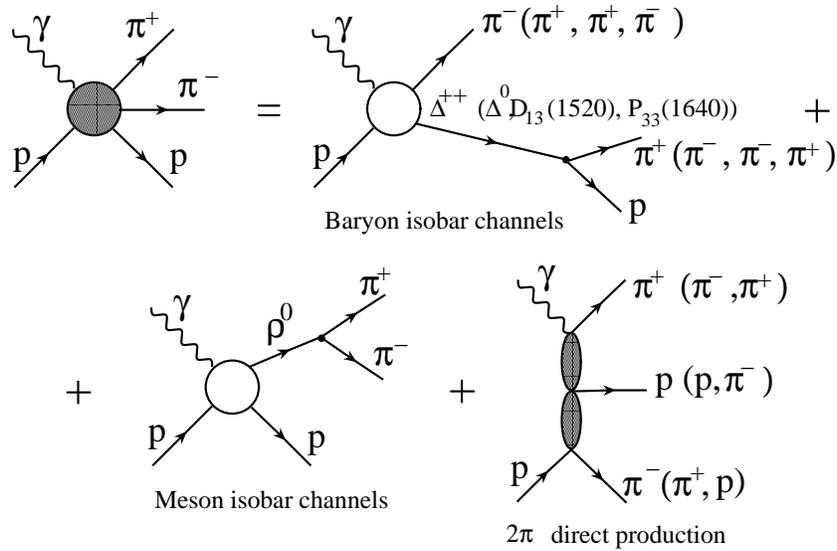,width=12cm}
\end{center}
\caption{The mechanisms of the JM model.}
\label{mech_main}
\end{figure*}

%\newpage
A comprehensive data set on 2$\pi$ single-differential and 
fully-integrated electroproduction cross sections measured with CLAS
has enabled us to establish the presence and strengths of the essential
$p\pi^+\pi^-$ electroproduction mechanisms. This was achieved within
the framework of a phenomenological model that has been developed over 
the past several years by the  Jefferson Laboratory - Moscow
State University collaboration (JM)
\cite{Ri00,Mo01,Mo03,Az06,Mo06,Shv07,Mo06-1,Mo08m,Mo07} for the analysis of
2$\pi$ photo- and electroproduction data. In this approach the resonant
part of the amplitudes is isolated and the $Q^2$ evolution of the
individual electrocoupling parameters of the contributing nucleon
resonances are determined from a simultaneous fit to all measured
observables.

The mechanisms of 2$\pi$ electroproduction incorporated into the JM
model are illustrated in Fig.~\ref{mech_main}. The full amplitudes
are described by superposition of the $\pi^{-} \Delta^{++}$, $\pi^{+}
\Delta^{0}$, $\rho p$, $\pi^{+} D^0_{13}(1520)$, $\pi^{+}
F^0_{15}(1685)$, and  $\pi^{-} P^{++}_{33}(1600)$ isobar channels
and the direct 2$\pi$ production mechanisms, where the $\pi^{+} \pi^{-} p$ 
final state is directly created without the formation of unstable
hadrons in the intermediate states. Nucleon resonances contribute to 
the baryon $\pi \Delta$ and meson $\rho p$ isobar
channels. The respective resonant amplitudes are evaluated in a
Breit-Wigner ansatz, as described in \cite{Ri00}. We included all 
well-established resonance states with hadronic decays to $N\pi\pi$ and an
additional $3/2^{+}$(1720) candidate state. Evidence for this
candidate state was found in the analysis of the CLAS 2$\pi$
electroproduction data \cite{Ri03}.

The $\pi \Delta$ isobar channels are strongest contributors to the
2$\pi$ electroproduction up to an invariant mass of $W \sim 2.0$~GeV. 
They have been clearly identified in the $\pi^{+} p$ and $\pi^{-} p$
1-fold differential mass distribution cross sections. The non-resonant
$\pi \Delta$  amplitudes  are calculated from the well established
Reggeized Born terms \cite{So71,Ri00,Mo07}. The initial and final
state interactions are described by an effective
absorptive-approximation~\cite{Ri00}. An additional contact term has
been introduced in~\cite{Mo06,Mo06-1,Mo07} to account phenomenologically for 
all remaining possible production mechanisms through the $\pi \Delta$
intermediate state channels, as well as for remaining FSI effects. The
parametrization for these amplitudes can be found in \cite{Mo07}.

The $\rho p$ isobar  channel becomes visible in the data at $W > 1.65$~GeV
with significant resonant contributions for  $W < 2.0$~GeV. Here the
non-resonant amplitudes are estimated by a diffractive ansatz, that
has been modified in order to reproduce experimental data in the near
and sub-threshold regions \cite{Shv07}. 

 The contributions from $\pi^{+} D_{13}^{0}(1520)$, $\pi^{+}
F_{15}^{0}(1685)$, $\pi^{-} P_{33}^{++}(1640)$ isobar channels are
seen in $\pi^{-} p$  and $\pi^{+} p$ mass distributions at $W >
1.65$~GeV. The $\pi^{+} D_{13}^{0}(1520)$ amplitudes are derived from
the Born terms of the $\pi \Delta$ isobar channels by implementing an
additional $\gamma_{5}$-matrix that accounts for the opposite parity
of the $D_{13}(1520)$  with respect to the $\Delta$. The amplitudes of 
$\pi^{+} F_{15}^{0}(1685)$ and $\pi^{-} P_{33}^{++}(1640)$ isobar
channels are parametrized as Lorentz invariant contractions of the
initial and final particle spin-tensors and with effective propagators
for the intermediate state particles. The magnitudes of these
amplitudes are fit to the data.

\begin{figure*}[htp]
\begin{center}
\includegraphics[width=16cm]{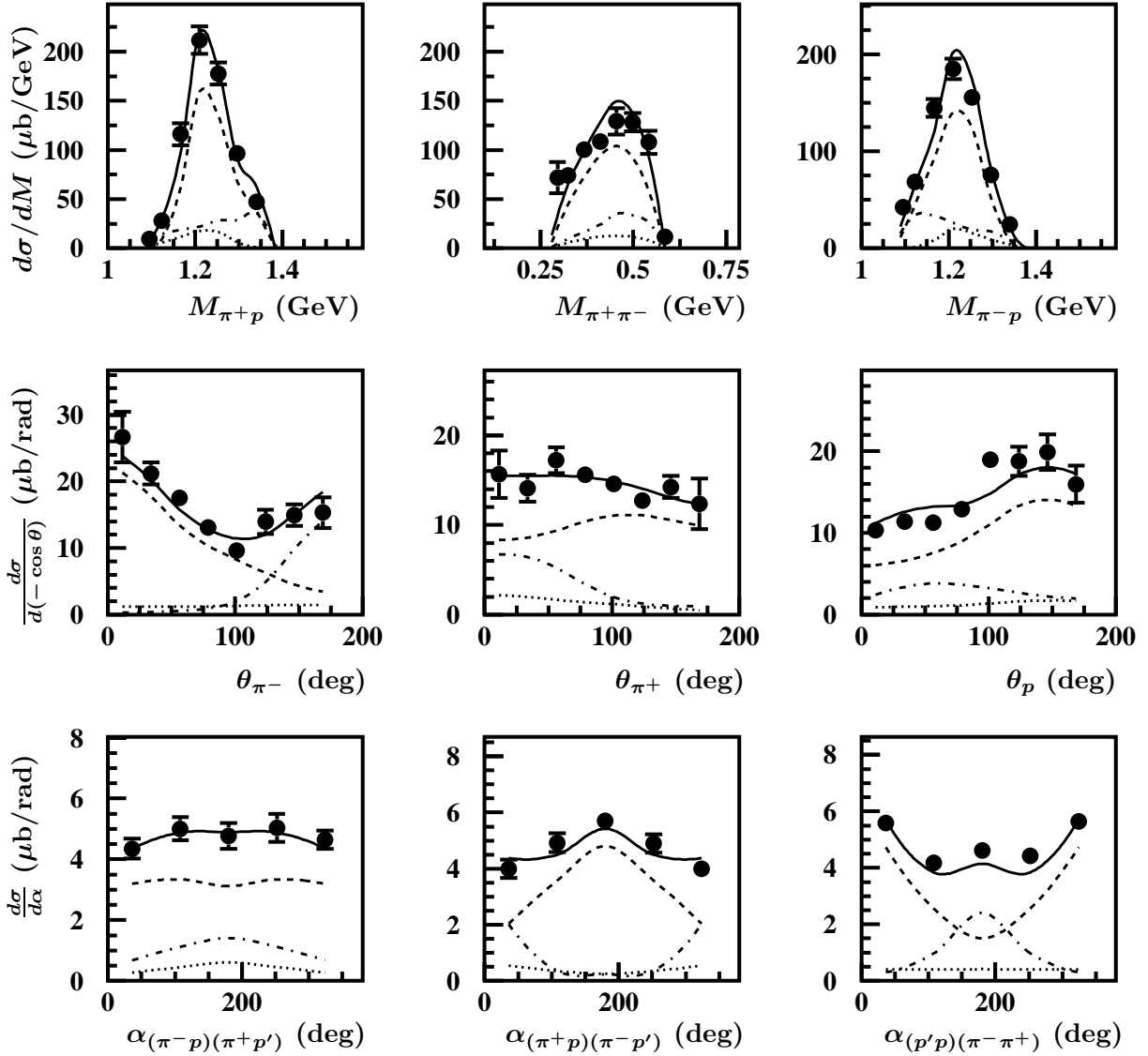}
\caption{Description of the CLAS charged double pion differential
  cross sections at $W = 1.51$~GeV and $Q^{2} = 0.425~GeV^{2}$
  within the framework of the JM model. Full calculations are shown by
  the solid lines. Contributions from $\pi^{-} \Delta^{++}$ and
  $\pi^{+} \Delta^{0}$ isobar channels are shown by the dashed and
  dotted lines, respectively, and contributions from the direct
  charged double pion production processes are shown by the dot-dashed
  lines.}
\label{9sectok}
\end{center}
\end{figure*}

\begin{figure*}[htp]
\begin{center}
\includegraphics[width=16cm]{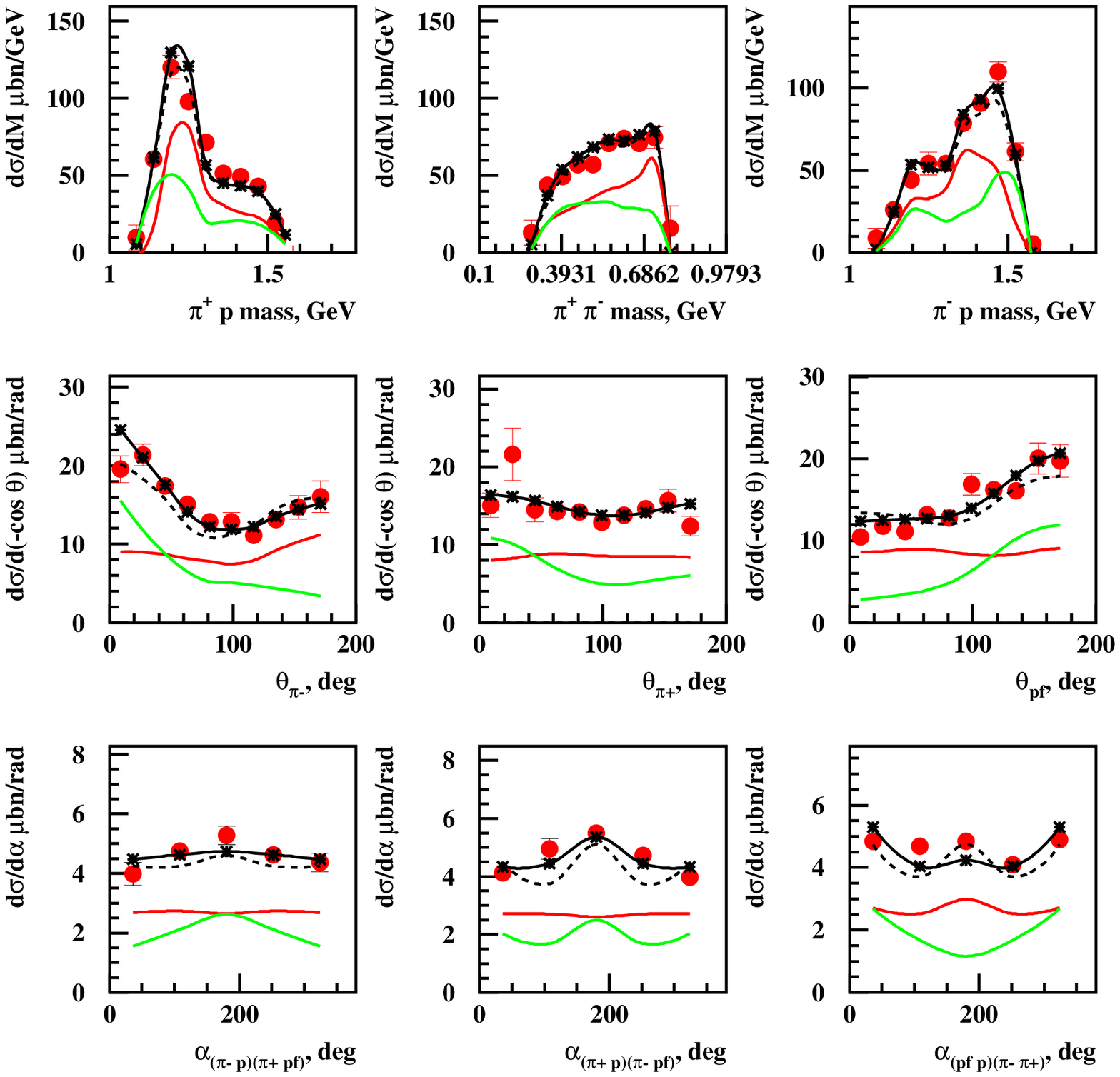}
\caption{Resonant (red lines) and non-resonant (green lines)
  contributions to the charged double pion differential cross sections
  at $W=1.71$~GeV and $Q^{2}=0.65$~GeV$^{2}$. The full JM
  calculation is shown by black lines, whereas the solid and dashed
  lines correspond to two different sets of $A_{1/2}$, $A_{3/2}$
  electrocoupling amplitudes for $3/2^{+}(1720)$ candidate state.}
\label{9secnstbck}
\end{center}
\end{figure*}

All isobar channels combined account for over 70\% of the charged
double pion production cross section in the nucleon resonance excitation region. 
The remaining part of cross sections stems from direct 2$\pi$
production processes, which are needed to describe backward strength
in the $\pi^{-}$ angular distributions and constrained by the 
$\pi^{+}$ and proton angular distributions~(see
Fig.~\ref{9sectok}). The strengths of the direct 2$\pi$ production
mechanisms, shown in bottom row of Fig.~\ref{mech_main}, have been
fitted to the CLAS cross section data \cite{Ri03,Fe07,Fe007,Fe07a} and can be found in
\cite{Mo07}.

\begin{figure}[ht]
%\begin{center}
%\hspace{-4cm}
\epsfig{file=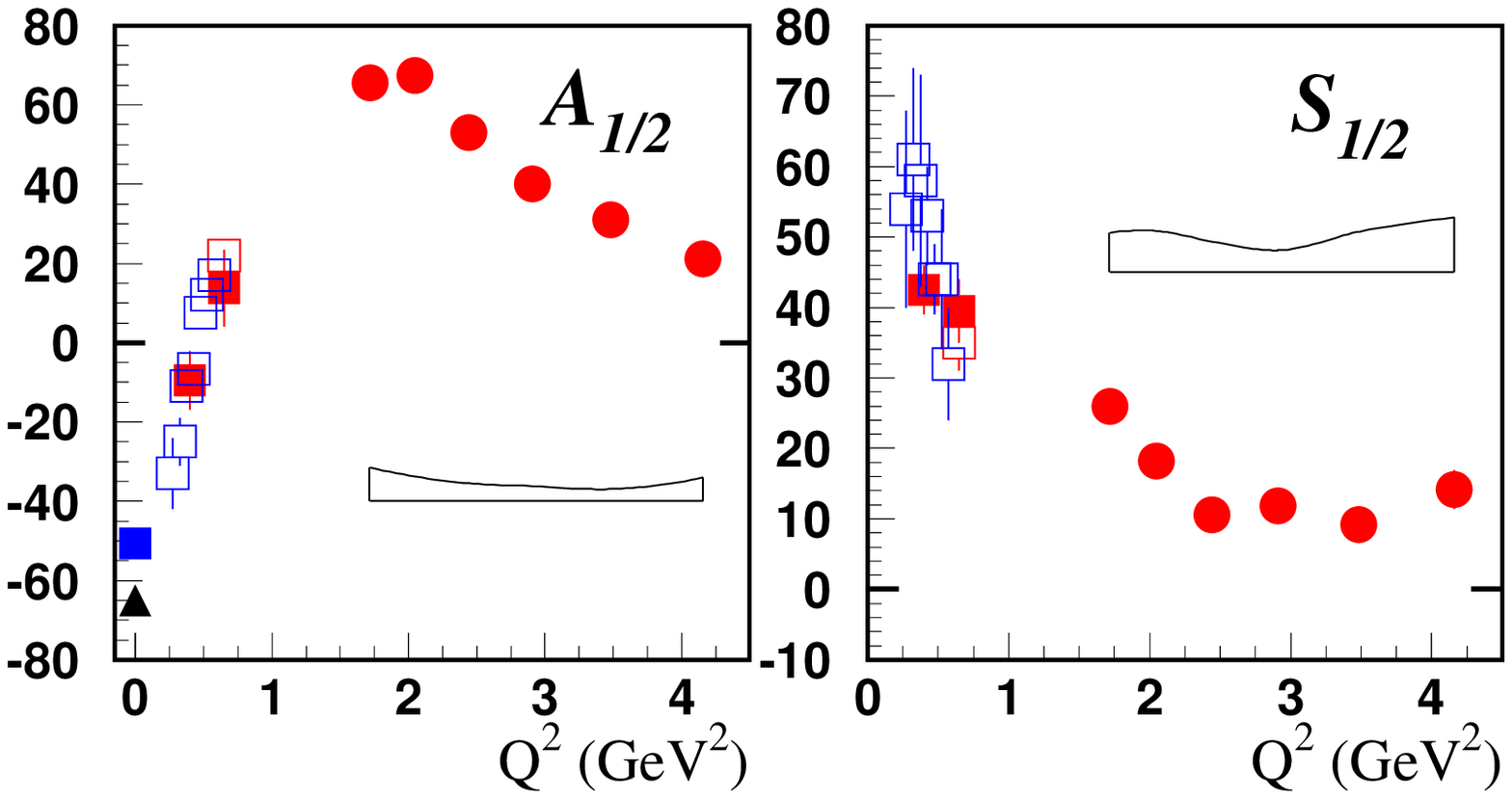,width=10.cm}
\epsfig{file=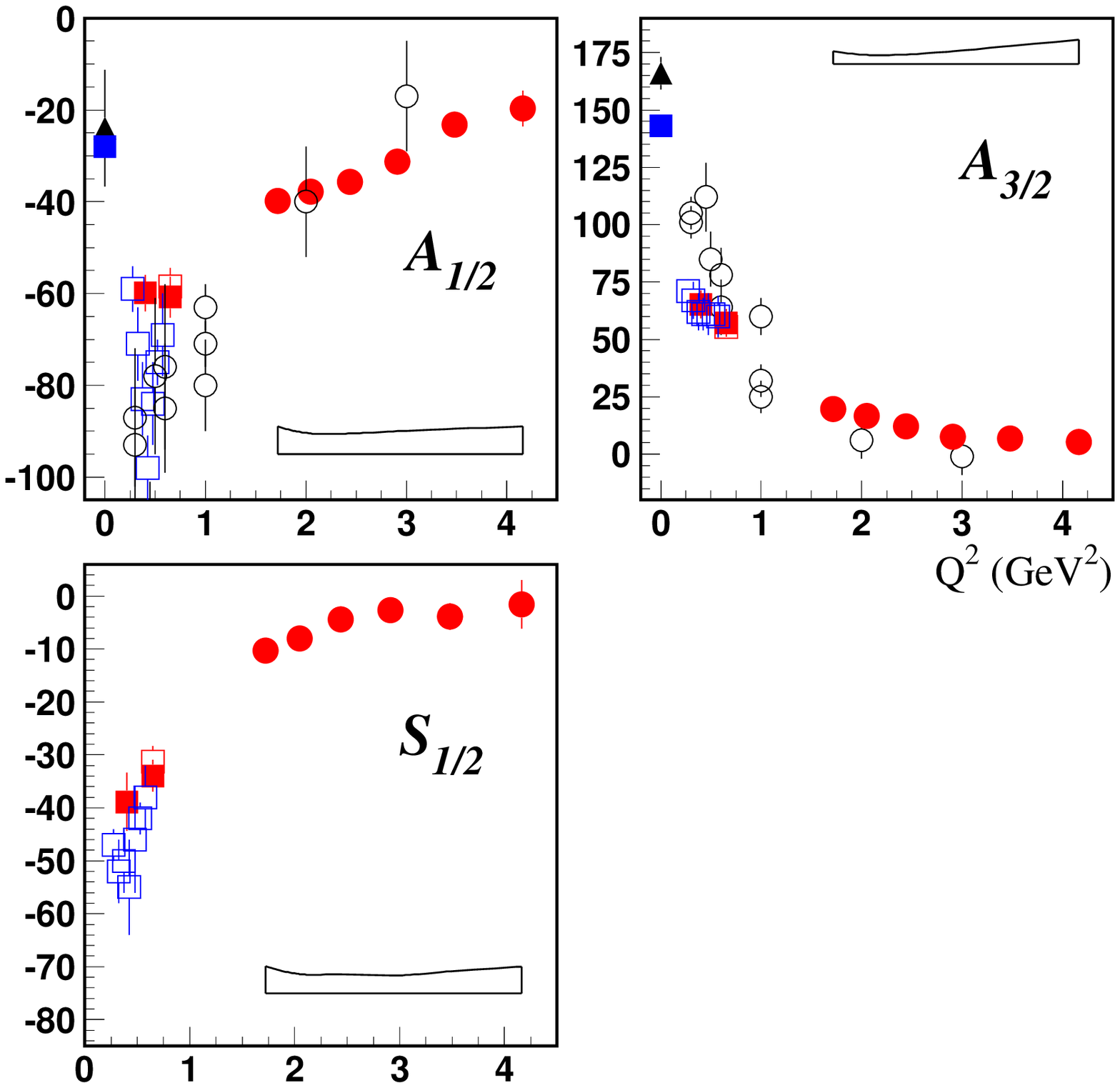,width=10.cm}
%\end{center}
\caption{Electrocoupling parameters of the $P_{11}(1440)$ (top row)  and 
$D_{13}(1520)$ (middle and bottom rows)  
 on the proton in units of $10^{-3}$~GeV$^{-1/2}$. CLAS
  results~\cite{Az05d,Az06,Az08g,Bu08c} of the 1$\pi$ production data are 
  represented by the red
  circles and squares. Open squares are for the combined
  analysis of the 1$\pi$ and 2$\pi$ channel~\cite{Az06}. 
  Blue open squares are the results of the 2$\pi$ data~\cite{Mo08m,Mo09ch} 
  at low $Q^{2}$. World data from 1$\pi^\circ$ electroproduction, 
  available before CLAS, are represented by black open circles. }
\label{p11d13}
\end{figure}

Within the framework of the JM approach we achieved a good description
of the 2$\pi$ data over the entire kinematic range covered by the
measurements.  As a typical example, the model description of the nine
1-fold differential cross sections at $W = 1.51$~GeV and $Q^{2} =
0.425$~GeV$^{2}$ are presented in Fig.~\ref{9sectok} together with
the contributing mechanisms. The different mechanism result in
qualitatively different shapes of their respective contributions to 
various observables. The successful simultaneous description of the nine
1-fold differential cross sections enables us to identify the 
essential contributing processes and to access their dynamics at the
phenomenological level. The extension of this approach to higher 
masses and higher $Q^2$ using data obtained at 6 GeV beam energy is 
currently underway. It will also provide new information on the 
mechanism that may be relevant for the phenomenological analysis 
of 2$\pi$ data at the 12 GeV upgrade. 

The amplitudes of non-resonant mechanisms derived from fitting the JM parameters to these data
may also be used as input for $N^{*}$ studies based on the global
multi-channel analysis in a fully coupled-channel approach that is
currently being developed at EBAC~\cite{Lee06,Lee07a,Lee07}. 

The separation of resonant and non-resonant contributions based on the
JM model parameters are shown in Fig.~\ref{9secnstbck}. Resonant and 
non-resonant parts have
qualitatively different shapes in all observables. This allows us to
isolate the resonant contributions and to extract the $N^{*}$
electrocoupling parameters.

\vspace{0.1in}
\begin{center}
{\bf N$^{*}$ electrocoupling parameters from single and double meson 
electroproduction}
\end{center}
\vspace{0.1in}

The CLAS data have enabled us for the first time to determine the
$P_{11}(1440)$, $D_{13}(1520)$ and $S_{11}(1535)$ electrocoupling
parameters over a wide range of photon virtualities by analyzing the
two major exclusive channels: 1$\pi$ and 2$\pi$ electroproduction.
These analyses have been carried out within the
framework of the approaches described above. The
electrocoupling parameters of the $P_{11}(1440)$ and $D_{13}(1520)$
states are shown in
Fig.~\ref{p11d13}. The agreement of the results
obtained from the analyses of 1$\pi$ and 2$\pi$ channels is highly 
significant since the 1$\pi$ and 2$\pi$ meson electroproduction
channels have completely different non-resonant amplitudes. The
successful description of the large body of data on
1$\pi$ and 2$\pi$ electroproduction with almost the same values 
for the $P_{11}(1440)$ and $D_{13}(1520)$ electrocoupling parameters,
shows the capability of the analyses methods to provide a
reasonable evaluation of the resonance parameters.

 The resonant part increases relative to the
non-resonant part with $W$ and $Q^{2}$. At $W>1.65$~GeV and 
$Q^2$ $>$ 0.5 $GeV^2$ it becomes the
largest contribution (see Fig.~\ref{9secnstbck}).  The 2$\pi$
electroproduction channel hence offers the best opportunity to
study higher-lying resonances ($W > 1.65$~GeV). The majority of these states 
decay dominantly by 2$\pi$ emission. Therefore, the combined analysis of 1$\pi$ and
2$\pi$ electroproduction opens the realistic possibility of accessing the 
electrocoupling parameters of the majority of excited states on the proton.

The results obtained from the 1$\pi$ and 2$\pi$ data represent reasonable, 
initial estimates of the
$Q^2$ evolution of the $N^{*}$ electrocoupling parameters. This
information will be checked and improved in a global and complete
coupled-channel analysis of major meson electroproduction channels 
that incorporates the amplitudes of 
non-resonant electroproduction mechanisms extracted from the CLAS data using the
phenomenological models described above. This program requires a
joint effort between Hall B and EBAC at Jefferson Lab.

%\end{document}

%\newpage
\section{      Status  of the Exited Baryons Analysis Center }

%\documentclass[preprint,aps,tightenlines,showpacs,superscriptaddress]{revtex4}
%\documentclass[prl,eqsecnum,twocolumn,floats,aps,showpacs,superscriptaddress]{revtex4}
%\usepackage[dvips]{graphicx}
%\usepackage{dcolumn}
%\usepackage{bm}
%\usepackage{epsfig}

%\usepackage{showlabels}
%\renewcommand{\case}{\frac}

%\begin{center}
%{\Large\bf Research Program at Excited Baryon Analysis Center}
%{\large T.-S. Harry Lee for EBAC collaboration}
%\end{center}

\vspace{0.1in}
\begin{center}
{\bf EBAC Strategy}
\end{center}
\vspace{0.1in}

The objective of EBAC is  more than just performing the
partial-wave analysis of the world data of
the $\pi N$, $\gamma N$ and $N(e,e^\prime)$ reactions. We  not only want to
 extract the $N^*$ parameters,  but also want to map out the
quark-gluon substructure of $N^*$ states. Thus
it requires a full dynamical coupled-channels 
analysis \cite{Lee06} which accounts for both 
the unitarity conditions and the reaction mechanisms at short distances.
The channels included in the current analysis
are two-particles $\gamma^* N, \pi N, \eta N, 
K\Lambda, K\Sigma, \omega N$ states and the crucial three-particle
$\pi\pi N$ state which has $\pi\Delta, \rho N, \sigma N$ resonant components.

The resonance parameters are extracted \cite{ssl08}
from the poles on the unphysical sheets of the complex-energy plane. 
Within the Hamiltonian formulation of the
constructed coupled-channel model, this method is capable of 
distinguishing the resonances
originating either from the meson-baryon attractive forces 
or from the excitations of the quark-gluon degrees of freedom  of the nucleon.
Clearly, this approach  is essential for
interpreting the extracted $N^*$ parameters in terms
of the predictions from hadron models and LQCD.

\begin{figure}[h]
\centering
\includegraphics[clip,width=10cm]{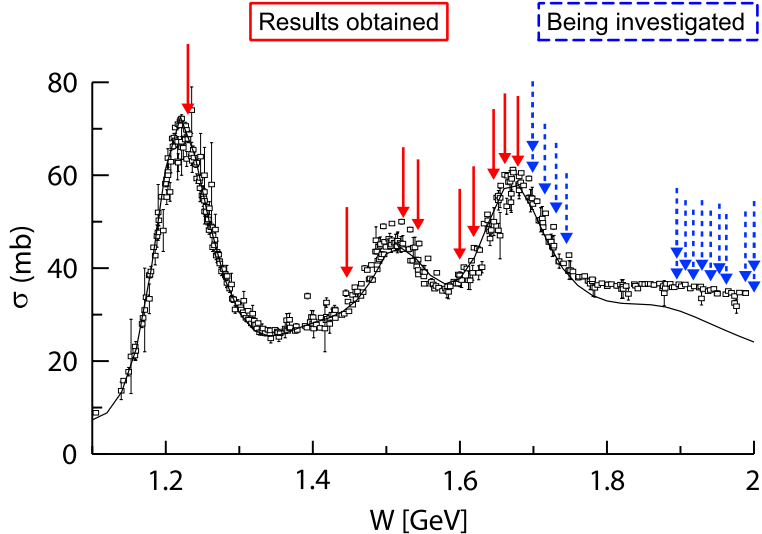}
\caption{The  $N^*$ positions listed by the Particle Data Group are
identified with the $\pi^- p$ total cross sections. The solid curve
is from  EBAC's model \cite{jlms}.}
\label{fig:ebac-status}
\end{figure}

\vspace{0.1in}
\begin{center}
{\bf Status of EBAC Analysis}
\end{center}
\vspace{0.1in}

\begin{figure}[h]
\centering
\includegraphics[clip,width=7.cm]{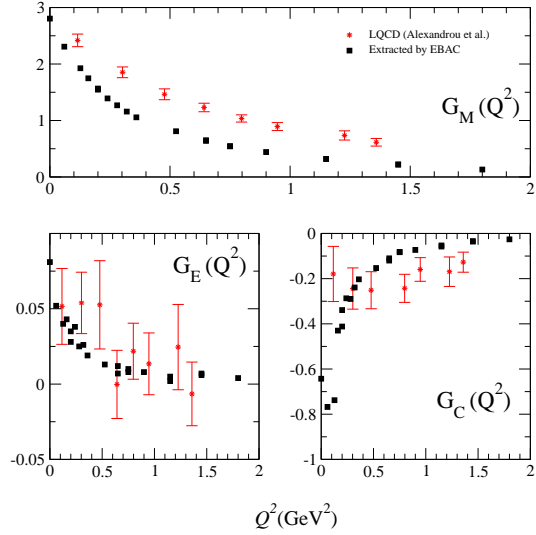}
\caption{The
$N$-$\Delta(1232)$ transition form factors $G_M$, $G_E$, and $G_C$ vs
$Q^2$. Empirical values (solid squares) are extracted by EBAC from world
data within a dynamical model. The LQCD results are from
Ref.~\cite{lqcd-dinna}.}
\label{fig:lattice-result}
\end{figure}

\begin{figure}[h]
\centering
\includegraphics[clip,width=4.5cm]{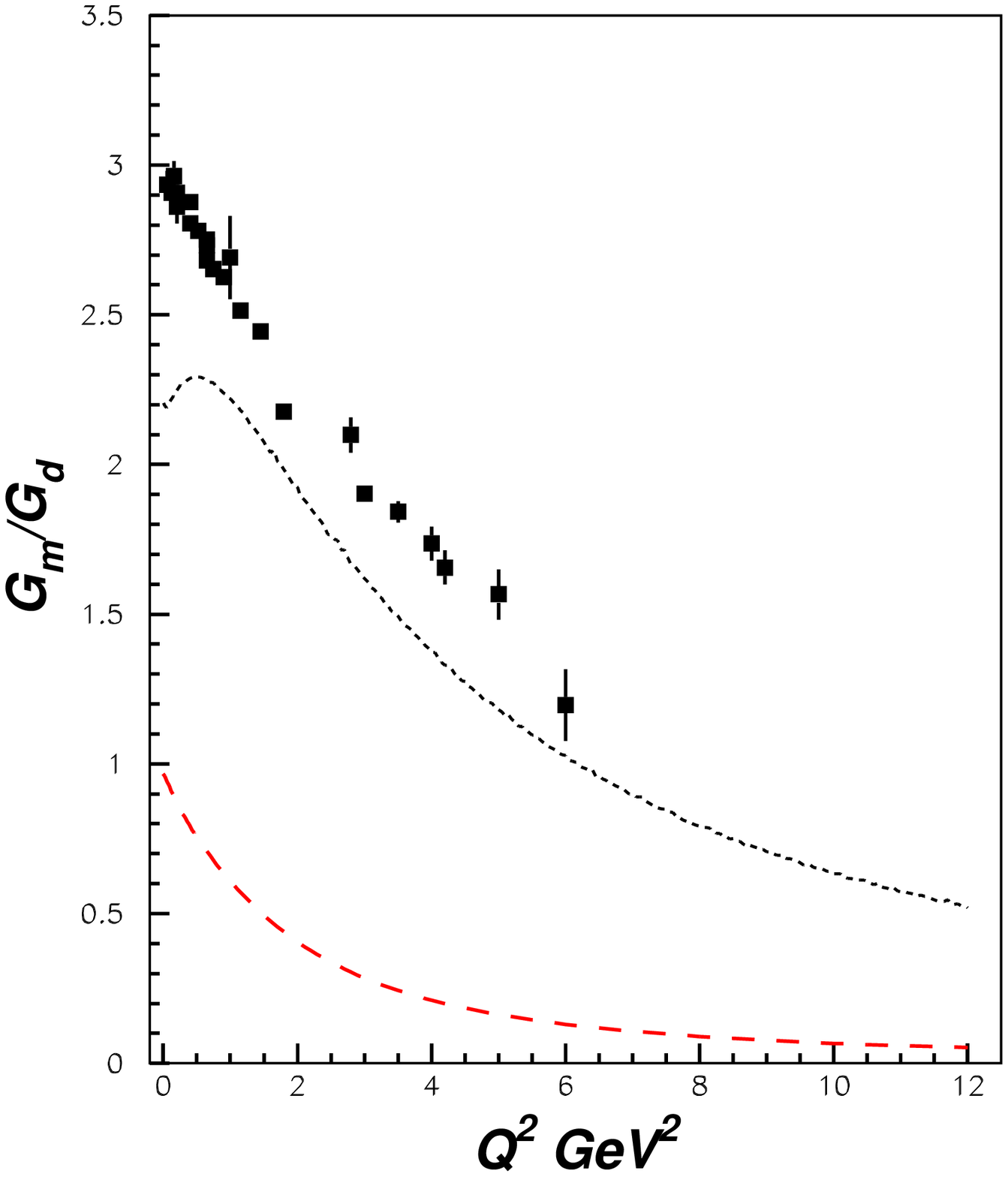}
\includegraphics[clip,width=4.5cm]{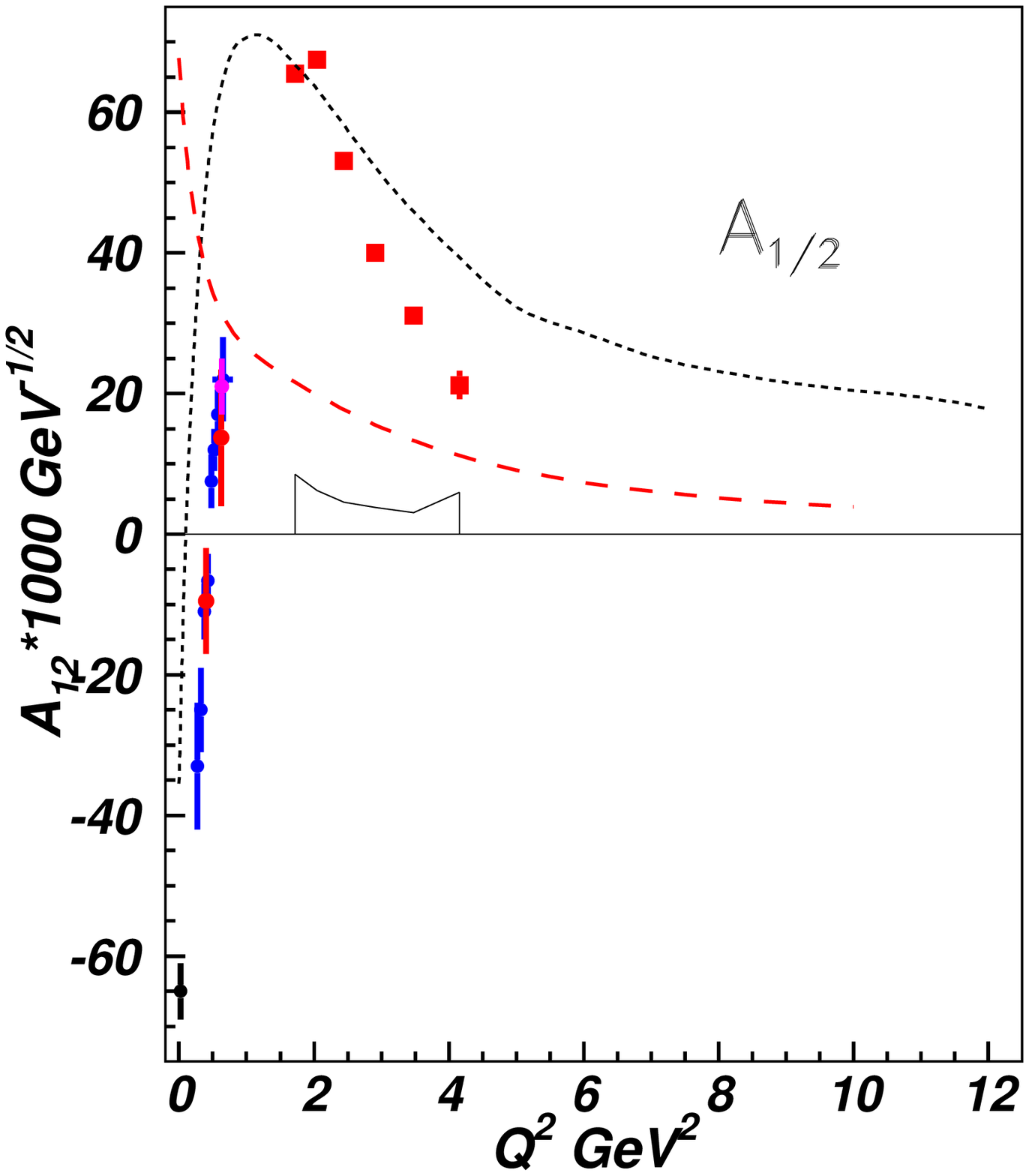}
\includegraphics[clip,width=4.5cm]{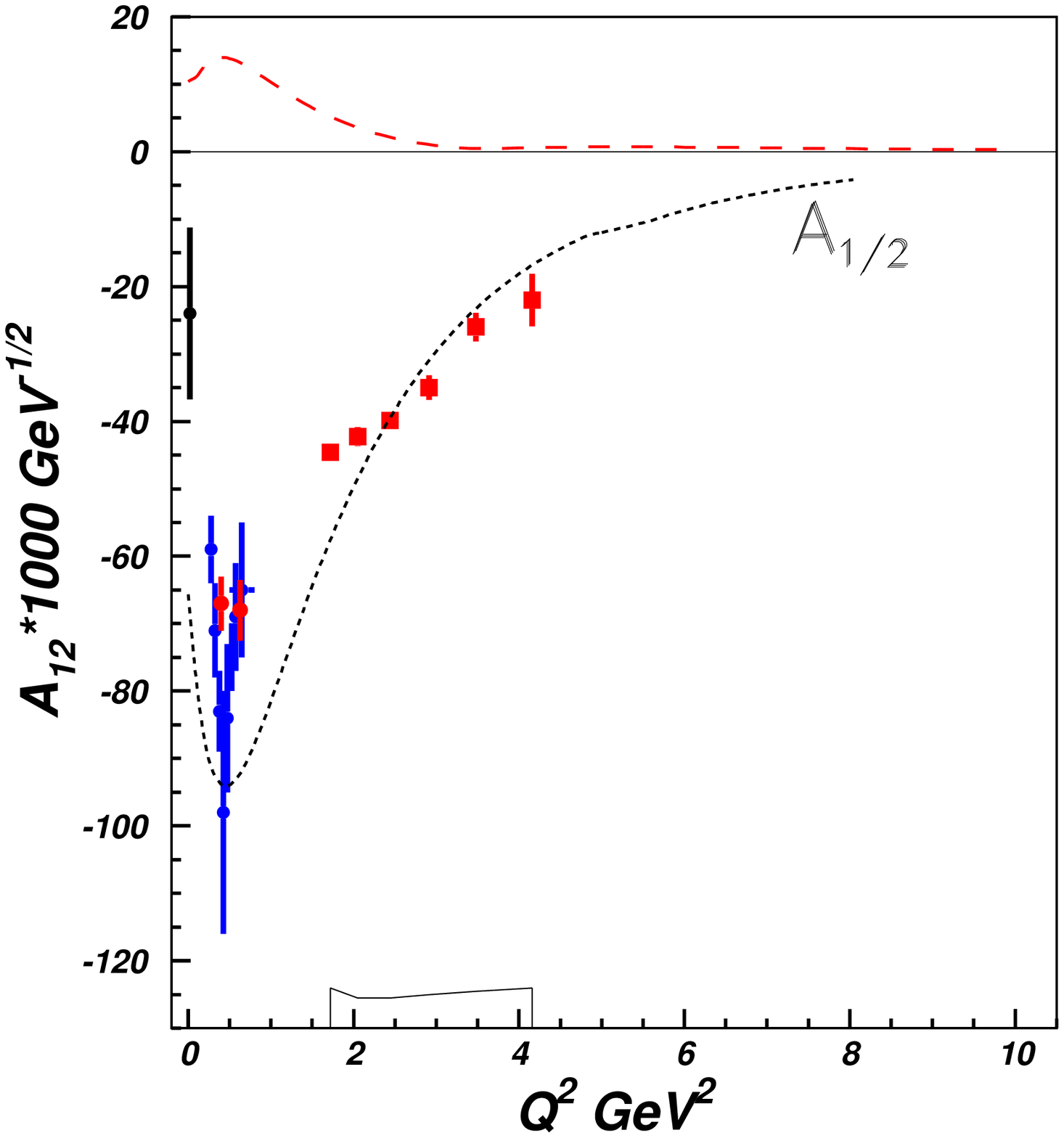}
\caption{The $N$-$N^*$ transition form factors and helicity amplitudes. Left panel: Magnetic form
factor for the $N$-$\Delta(1232)$ transition normalized to the dipole form
factor. Center panel: Transition helicity amplitude $A_{1/2}$ for
$N$-$N^{*}(1440)$. Right panel: Transition helicity amplitude $A_{1/2}$ for
$N$-$N^{*}(1520)$.  The results from CLAS/world experimental data analyses
are shown by the data points~\cite{Bu08c,Burk05a,Lee04a,Mo08m,Az08}.  The red
and blue symbols are the results from analyses of 1$\pi$ and 2$\pi$ exclusive
channels, respectively. The curves are from Constituent Quark Model
calculations
(dotted), described in section VI, and from meson-baryon dressing contributions predicted by the
EBAC (dashed).}
\label{fig:cqm-result}
\end{figure}

The analysis of $\pi N, \gamma^*N \rightarrow \pi N, \eta N, \pi\pi N$ has 
been performed \cite{jlss,jlms,jlmss,djlss,kjlms}. The resonance parameters of
the low-lying  $N^*$ states
with masses below about 1.7 GeV (red arrows in Fig.\ref{fig:ebac-status})
 have been extracted.
In Fig.\ref{fig:lattice-result}, we show the comparison of the
$N$-$\Delta(1232)$ transition form factors $G_M$, $G_E$, and $G_C$ vs
$Q^2$  extracted by EBAC from world
data with the LQCD results from
Ref.~\cite{lqcd-dinna}. In Fig.\ref{fig:cqm-result}, we show that
the discrepancies between the 
$\gamma N \rightarrow \Delta (1232),
N^*(1440), N^*(1520)$ form factors predicted by the
constituent quark models (dotted curves) and the empirical values from CLAS
collaboration could be accounted for by including the 
meson-baryon dressing (meson cloud) 
effects (red dashed curves) predicted by the EBAC
collaboration.

 The higher mass $N^*$ states ( dashed arrows in Fig.\ref{fig:ebac-status}), 
suggested by Particle Data Group, are still being investigated.
The main task is to include $K\Lambda$, $K\Sigma$, and $\omega N$ 
channels in the analysis. Furthermore an effort has been devoted to
recover the old data of $\pi N \rightarrow \pi\pi N$ reactions which
are essential in pinnig  down the higher mass $N^*$ states. New data from
new hadron facilities, such as JPARC in Japan,  perhaps will be essential
 in making  conclusive determinations
of these ``elusive" $N^*$ states.

The constructed coupled-channel model
can be  extended to include other resonant channels, such
as the $\pi N^*( D_{13}, 1520),\pi N^*( F_{15}, 1680)
,\pi \Delta ( P_{33},1600)$ 
channels suggested by the  CLAS collaboration,
if necessary.
We are also investigating how the K-matrix models, used in the data analysis
by the CLAS collaboration
(described in section VII), the Mainz group, and the Bonn group, 
can be related to 
EBAC's dynamical approach.
This will then further strengthen the  theory-experiment joint effort in
extracting the parameters of the higher mass $N^*$'s indicated in
Fig.\ref{fig:ebac-status}.

With the data from 12 GeV upgrade, the EBAC analysis needs to be extended
to account for reaction mechanisms at high momentum transfer.
It is necessary to describe the nonresonant mechanisms directly in terms
of quark-gluon degrees of freedom. The results from DSE models,
described in section III,  will be used to make progress in this direction.

%\begin{figure}[h]
%\centering
%\includegraphics[clip,width=15cm]{STATUS_EBAC/ebac-plan.eps}
%\caption{EBAC Strategy}
%\label{fig:ebac-plan}
%\end{figure}
%\newpage

%%%%%%%%%%%%%%%%%%%%%%%%%%%%%%%%%%%%%%%%%%%%%%%%%%%%%%%%%%%%%%%%%%%%%%%%%%%
%%%%%%%%%%%%%%%%%%%%%%%%%%%%%%%%%%%%%%%%%%%%%%%%%%%%%%%%%%%%%%%%%%%%%%%%%%%

\clearpage
%References

%\end{document}

%\newpage
\section*{Acknowledgement}
The authors are thankful to Dr. L.~Elouadrhiri for useful discussions and
important support. We express our gratitude to the Staff Services group 
at Jefferson Lab, especially to Mrs. S.~Schatzel, the Electromagnetic 
$N$--$N^*$ Transition Form Factors 
 Workshop Secretary, for her invaluable  administrative support of this very
 successful scientific meeting.

\newpage

%\appendix

%\section*{Bibliography} 

%\section{The Program of the Electromagnetic $N$--$N^*$ Transition Form Factors Workshop}
%\input{workshop_prog.tex}

%\newpage
%\section{Participants of the Electromagnetic $N$--$N^*$ Transition Form Factors Workshop}

\end{document}